\documentclass[submission,Phys]{SciPost}
\usepackage{lmodern}

\usepackage[T1]{fontenc}
\usepackage{geometry}
\usepackage{color}
\usepackage[english]{babel}
\usepackage{float}
\usepackage{amsmath}
\usepackage{amssymb}
\usepackage{graphicx}
\usepackage[numbers,sort&compress]{natbib}

\makeatletter
\@ifundefined{date}{}{\date{}}
\makeatother

\begin{document}
\title{Entanglement Measures in a Nonequilibrium Steady State: Exact Results
in One Dimension}
\author{Shachar Fraenkel{*} and Moshe Goldstein\\
\emph{\normalsize{}Raymond and Beverly Sackler School of Physics and
Astronomy, Tel-Aviv University,}\\
\emph{\normalsize{}Tel Aviv 6997801, Israel}\\
{\normalsize{}{*}shacharf@mail.tau.ac.il}}
\maketitle

\section*{Abstract}

\textbf{Entanglement plays a prominent role in the study of condensed
matter many-body systems: Entanglement measures not only quantify
the possible use of these systems in quantum information protocols,
but also shed light on their physics. However, exact analytical results
remain scarce, especially for systems out of equilibrium. In this
work we examine a paradigmatic one-dimensional fermionic system that
consists of a uniform tight-binding chain with an arbitrary scattering
region near its center, which is subject to a DC bias voltage at zero
temperature. The system is thus held in a current-carrying nonequilibrium
steady state, which can nevertheless be described by a pure quantum
state. Using a generalization of the Fisher-Hartwig conjecture, we
present an exact calculation of the bipartite entanglement entropy
of a subsystem with its complement, and show that the scaling of entanglement
with the length of the subsystem is highly unusual, containing both
a volume-law linear term and a logarithmic term. The linear term is
related to imperfect transmission due to scattering, and provides
a generalization of the Levitov-Lesovik full counting statistics formula.
The logarithmic term arises from the Fermi discontinuities in the
distribution function. Our analysis also produces an exact expression
for the particle-number-resolved entanglement. We find that although
to leading order entanglement equipartition applies, the first term
breaking it grows with the size of the subsystem, a novel behavior
not observed in previously studied systems. We apply our general results
to a concrete model of a tight-binding chain with a single impurity
site, and show that the analytical expressions are in good agreement
with numerical calculations. The analytical results are further generalized
to accommodate the case of multiple scattering regions.}

\section{Introduction\label{sec:Introduction}}

Soon after the nascence of quantum mechanics, entanglement was recognized
as a unique attribute of quantum systems. Yet only over the past few
decades has its consequential nature truly been acknowledged, in particular
in the realm of many-body physics~\citep{RevModPhys.80.517,RevModPhys.81.865,LAFLORENCIE20161}.
Entanglement is now routinely harnessed to detect quantum phase transitions,
both in and out of equilibrium~\citep{Osterloh2002,PhysRevA.66.032110,PhysRevLett.90.227902,Calabrese_2004,PhysRevB.82.174411,RevModPhys.91.021001};
to characterize long-range correlations in contexts such as dynamics~\citep{Calabrese_2005,Chiara_2006,Peschel_2009,PhysRevLett.111.127205,PhysRevB.90.174202}
or topological order~\citep{PhysRevLett.96.110404,PhysRevLett.96.110405,PhysRevLett.101.010504};
and to analyze the capabilities and limitations of simulation methods~\citep{PhysRevLett.91.147902,PhysRevLett.93.227205,doi:10.1080/14789940801912366}.

The prospect of promoting our understanding of nonequilibrium quantum
systems through the investigation of their entanglement properties
is especially intriguing. The study of quantum many-body phenomena
out of equilibrium has shown promising progress in recent years, propelled
by the development of suitable quantum simulation platforms in cold
atom systems~\citep{Cheneau2012,Trotzky2012,Gring1318,Langen2013,Schreiber842}.
But while the theoretical understanding of quantum many-body nonequilibrium
has substantially progressed~\citep{RevModPhys.83.863,Gogolin_2016},
rigorous analytical results are still rare. Entanglement measures
have established a route for producing such results~\citep{PhysRevA.78.010306,Eisler_2012,PhysRevA.89.032321,PhysRevB.96.054302,10.21468/SciPostPhys.8.3.036,Gruber_2020,Collura_2014,Eisler_2014,HOOGEVEEN201578,alba2021unbounded},
a route which this work seeks to advance.

Entanglement entropy~\citep{RevModPhys.80.517,RevModPhys.81.865}
is a measure usually employed to quantify entanglement within a many-body
system in a pure state, represented by a density matrix $\rho$. Given
a bipartition of the total system into subsystems $A$ and $B$, one
obtains the reduced density matrix (RDM) of subsystem $A$ by tracing
out the degrees of freedom associated with subsystem $B$, $\rho_{A}={\rm Tr}_{B}\left[\rho\right]$.
The von-Neumann entanglement entropy (vNEE) is then defined as 
\begin{equation}
{\cal S}=-{\rm Tr}\left[\rho_{A}\ln\rho_{A}\right].
\end{equation}
Additionally, we denote the $n$th moment of the RDM as
\begin{equation}
Z_{n}={\rm Tr}\left[\rho_{A}^{n}\right],
\end{equation}
and refer to it as the R\'enyi moment of order $n$. Note that this
is slightly different than the R\'enyi entropy, $S_{n}=\frac{1}{1-n}\ln\left({\rm Tr}\left[\rho_{A}^{n}\right]\right)$.
The von-Neumann entropy and the R\'enyi moments are related through
\begin{equation}
{\cal S}=-\lim_{n\to1}\partial_{n}Z_{n}.\label{eq: vNEE relation to Renyi moments}
\end{equation}
Between these two measures of bipartite entanglement, the vNEE constitutes
the more rigorous one in and of itself~\citep{RevModPhys.80.517,RevModPhys.81.865}.
Nevertheless, R\'enyi entropies may be used to provide lower bounds
to the vNEE and to reconstruct the full entanglement spectrum~\citep{PhysRevLett.101.010504,PhysRevB.85.035409,PhysRevLett.109.020505},
and are also accessible to direct experimental measurement~\citep{PhysRevLett.109.020504,Islam2015}.

Of prime importance is the way in which the vNEE scales with the size
of the subsystem in question. This scaling law is considered to be
a significant classification criterion, distinguishing between typical
phases of condensed matter systems~\citep{LAFLORENCIE20161}. A prominent
example is the renowned area law, under which the vNEE scales linearly
with the area of the subsystem's boundary: ${\cal S}\sim cL^{d-1}$,
with $L$ being a typical linear dimension of the subsystem, $d$
being the spatial dimension and $c$ being a constant~\citep{RevModPhys.82.277}.
The area law generally applies to ground states of gapped local Hamiltonians~\citep{PhysRevLett.94.060503,Hastings_2007},
as well as to excited eigenstates of many-body-localized systems~\citep{Bauer_2013,PhysRevLett.111.127201,PhysRevB.90.174202},
and has generated particular interest due to the fact that states
obeying the area law admit an efficient tensor-network representation~\citep{RevModPhys.77.259,PhysRevB.73.094423,Hastings_2007,PhysRevLett.100.030504,doi:10.1080/14789940801912366}.
Ground states of gapless systems (and specifically critical systems)
with a finite and sharp Fermi surface tend to violate the area law
by a logarithmic correction, ${\cal S}\sim cL^{d-1}\ln L$~\citep{PhysRevLett.90.227902,Calabrese_2004,PhysRevLett.96.010404,PhysRevLett.96.100503}.
On the other hand, a volume-law scaling -- i.e., an extensive scaling
of the vNEE, ${\cal S}\sim cL^{d}$ -- is widely observed in highly-excited
states of local thermalizing Hamiltonians~\citep{Alba_2009,doi:10.1146/annurev-conmatphys-031214-014726,PhysRevB.91.155123,doi:10.1080/00018732.2016.1198134}
and in ground states of non-local Hamiltonians~\citep{Alba_2009}.

An additional tool in the analysis of many-body entanglement, that
has lately come into increased awareness, is the symmetry- or charge-resolved
entanglement entropy~\citep{Laflorencie_2014,PhysRevLett.120.200602,PhysRevB.98.041106,PhysRevLett.121.150501,PhysRevA.100.022324}.
Given an additive quantity $Q=Q_{A}+Q_{B}$ that is globally conserved
in the total system (e.g., particle number), the RDM $\rho_{A}$ derived
for a pure eigenstate of the Hamiltonian turns out to be block-diagonal
with respect to the eigenbasis of $Q_{A}$, $\rho_{A}=\oplus_{Q_{A}}\rho_{A}^{\left(Q_{A}\right)}$.
This suggests that entanglement entropies may be calculated for each
block separately, thus resolving the total R\'enyi moments and vNEE
to sums over contributions from symmetry sectors:
\begin{align}
Z_{n} & =\sum_{Q_{A}}Z_{n}\left(Q_{A}\right)=\sum_{Q_{A}}{\rm Tr}\left[\left(\rho_{A}^{\left(Q_{A}\right)}\right)^{n}\right],\nonumber \\
{\cal S} & =\sum_{Q_{A}}{\cal S}\left(Q_{A}\right)=-\sum_{Q_{A}}{\rm Tr}\left[\rho_{A}^{\left(Q_{A}\right)}\ln\rho_{A}^{\left(Q_{A}\right)}\right].\label{eq: Symmetry resolved definition}
\end{align}
We note that the definition in Eq.~(\ref{eq: Symmetry resolved definition})
follows Refs.~\citep{Laflorencie_2014,PhysRevLett.120.200602}, while
in other works~\citep{PhysRevB.98.041106,PhysRevLett.121.150501,PhysRevA.100.022324}
each symmetry block is normalized by its trace before the resolved
moments and entropies are calculated, rendering them measures of entanglement
following a projection onto a symmetry sector. These normalized quantities
may be straightforwardly derived from their non-normalized counterparts
from Eq.~(\ref{eq: Symmetry resolved definition}) by relying on
the fact that $Z_{1}\left(Q_{A}\right)$ -- which is simply the charge
distribution in subsystem $A$ -- returns the required trace of each
block. Specifically, the post-projection vNEE is given by
\begin{equation}
\sigma\left(Q_{A}\right)=\ln\left(Z_{1}\left(Q_{A}\right)\right)+\frac{{\cal S}\left(Q_{A}\right)}{Z_{1}\left(Q_{A}\right)}.\label{eq: Definition for post measurement vNEE}
\end{equation}
The resolved quantities in Eq.~(\ref{eq: Symmetry resolved definition})
do not quantify entanglement when used alone, but are more readily
calculated, and can be also directly measured in experiments~\citep{PhysRevLett.120.050406,PhysRevA.97.023604,PhysRevLett.120.200602,PhysRevA.98.032302,PhysRevA.99.062309}.

Symmetry-resolved entanglement represents the internal structure that
symmetry imposes on the entanglement spectrum, thus embodying the
interplay between conservation laws and entanglement. It has been
recently investigated analytically and numerically in various systems,
in and out of equilibrium~\citep{PhysRevB.99.115429,PhysRevB.100.235146,Bonsignori_2019,Fraenkel_2020,10.21468/SciPostPhys.8.3.046,Capizzi_2020,PhysRevB.102.014455,Murciano_2020,PhysRevB.103.L041104,Calabrese_2020,PhysRevB.101.235169,PhysRevB.102.054302,Murciano2020,PhysRevLett.125.120502,Horvath2020,Bonsignori_2020,PhysRevB.102.235157,10.21468/SciPostPhys.10.3.054,zhao2020symmetryresolved,vitale2021symmetryresolved,murciano2021symmetry,horvath2021u1,neven2021symmetryresolved}.
A common behavior in these systems is entanglement equipartition~\citep{PhysRevB.98.041106,Bonsignori_2019,Fraenkel_2020,10.21468/SciPostPhys.8.3.046,PhysRevB.102.014455,Murciano_2020,Capizzi_2020,PhysRevB.103.L041104},
implying that, to leading order in $L$, the post-projection vNEE
$\sigma\left(Q_{A}\right)$ is constant across symmetry sectors. The
estimation of symmetry-resolved entanglement was shown to yield additional
valuable insights, e.g.~regarding topological phase transitions~\citep{Fraenkel_2020,PhysRevLett.125.120502,PhysRevB.102.235157}
and dissipation in noisy devices~\citep{vitale2021symmetryresolved}.

The main result of this work is the exact calculation of an unusual
entanglement scaling for the steady state of a one-dimensional fermionic
system out of equilibrium, using a generalization of the Fisher-Hartwig
conjecture~\citep{10.2307/23030524}. We study a model of a uniform
tight-binding chain containing an arbitrary scattering region at its
center, to which a DC bias voltage is applied at zero temperature,
thereby leading to a current-carrying steady state. This steady state
may be described by a pure eigenstate of the Hamiltonian, with different
distributions for scattering states incoming from the left and from
the right. This model is relevant to both electronic~\citep{datta_1995}
and cold atom~\citep{Husmann1498} systems.

We report that the subsystem entanglement entropy in this nonequilibrium
steady state exhibits a volume-law scaling accompanied by an additive
logarithmic correction. More precisely, we find that the vNEE of a
subsystem of length $L$, when located far enough from the scattering
region, obeys
\begin{equation}
{\cal S}\sim\left(-\underset{k_{-}}{\overset{k_{+}}{\int}}\frac{dk}{2\pi}\left[\left|t\left(k\right)\right|^{2}\ln\left(\left|t\left(k\right)\right|^{2}\right)+\left|r\left(k\right)\right|^{2}\ln\left(\left|r\left(k\right)\right|^{2}\right)\right]\right)L+{\cal C}_{{\rm log}}\ln L+{\cal C}_{{\rm const}}.\label{eq: main scaling law}
\end{equation}
Here $\left|t\left(k\right)\right|^{2}$ and $\left|r\left(k\right)\right|^{2}$
are the transmission and reflection factors (respectively) of the
scatterer for a plane wave with momentum $k$, $0\le k_{-}<k_{+}$
are the two different Fermi momenta for right- and left-propagating
fermions, and ${\cal C}_{{\rm log}},{\cal C}_{{\rm const}}$ are constants.
The extensive term of the vNEE is thus generated by momentum eigenstates
within the bias voltage window, and the contribution of each state
is equivalent to the classical mixture entropy of the corresponding
transmission probability. The logarithmic term in Eq.~(\ref{eq: main scaling law})
is a zero-temperature effect, that arises due to the sharp Fermi-Dirac
jumps in the distribution of the plane waves. The coefficient ${\cal C}_{{\rm log}}$
is therefore a function of the Fermi momenta, for which we provide
an exact expression as well. For ${\cal C}_{{\rm const}}$ we obtain
an approximate expression, justified under the assumption of a small
bias voltage.

Furthermore, we expand the result in Eq.~(\ref{eq: main scaling law})
by deriving the exact entanglement entropy asymptotics for the case
where the chain hosts multiple interspersed scattering regions, finding
a similar scaling law. This type of scaling has been previously encountered
in rather specific instances, for example in certain excited states,
in the ground state of 1D systems with long-range couplings~\citep{Alba_2009,Ares_2014},
and in the diffusive time-averaged state of a 1D interacting system~\citep{PhysRevB.100.134306,ippoliti2021fractal}.
We show how this entanglement scaling can arise generically within
a local 1D system in a time-independent state, generalizing previous
results in particular cases~\citep{PhysRevA.89.032321,PhysRevB.96.054302,Eisler_2012,10.21468/SciPostPhys.8.3.036,Gruber_2020}.

Our calculation also produces analytical results for the symmetry-resolved
entanglement, with respect to the total fermionic charge that is conserved
in the system. While the post-projection vNEE exhibits equipartition
to leading order as usual, we remarkably find that the first term
breaking equipartition may grow with the size of the subsystem, and
is anti-symmetric in $Q_{A}-\left\langle Q_{A}\right\rangle $, the
deviation of the charge in $A$ from its mean. To the best of our
knowledge, this is the first system found to display such behavior.

The paper is organized as follows: In Sec.~\ref{sec: The Physical Model}
we describe the general 1D nonequilibrium model, and derive expressions
for the elements of its two-site correlation matrix. In Sec.~\ref{sec:Analytical-Asymptotics}
we review the definition of the generating function that captures
all resolved and unresolved moments and entanglement entropies. By
employing a generalization of the Fisher-Hartwig conjecture, we analytically
derive the exact leading-order asymptotics of the generating function
for a large subsystem, along with an approximate subleading correction.
We use the result for the generating function to extract quantities
relating to charge statistics and the asymptotic scaling of the vNEE
in the nonequilibrium steady state. We also address charge-resolved
entanglement, and determine the form of the first term breaking entanglement
equipartition. In Sec.~\ref{sec: The-Single-Impurity} we apply our
general scheme to a specific model, where a single impurity site is
responsible for the scattering. We use this concrete example to demonstrate
central aspects of our results, and to show that the analytical calculation
generally compares favorably with numerics. Sec.~\ref{sec:Generalization-to-multiple}
details how the analytical results may be generalized to a subsystem
on a chain containing multiple but distant scattering regions. In
Sec.~\ref{sec:Conclusions-and-outlook} we discuss our main conclusions
and outline possible future directions. Appendix~\ref{sec:Detailed-derivations}
is dedicated to further technical details of the derivation. Appendix~\ref{sec:Appendix:-Additional-plots}
contains additional plots of the analytical results for the single
impurity model, and points out the generic features that apply to
the general model.

\section{Model\label{sec: The Physical Model}}

\begin{figure}[t]
\begin{centering}
\includegraphics[viewport=180bp 0bp 1100bp 640bp,clip,scale=0.44]{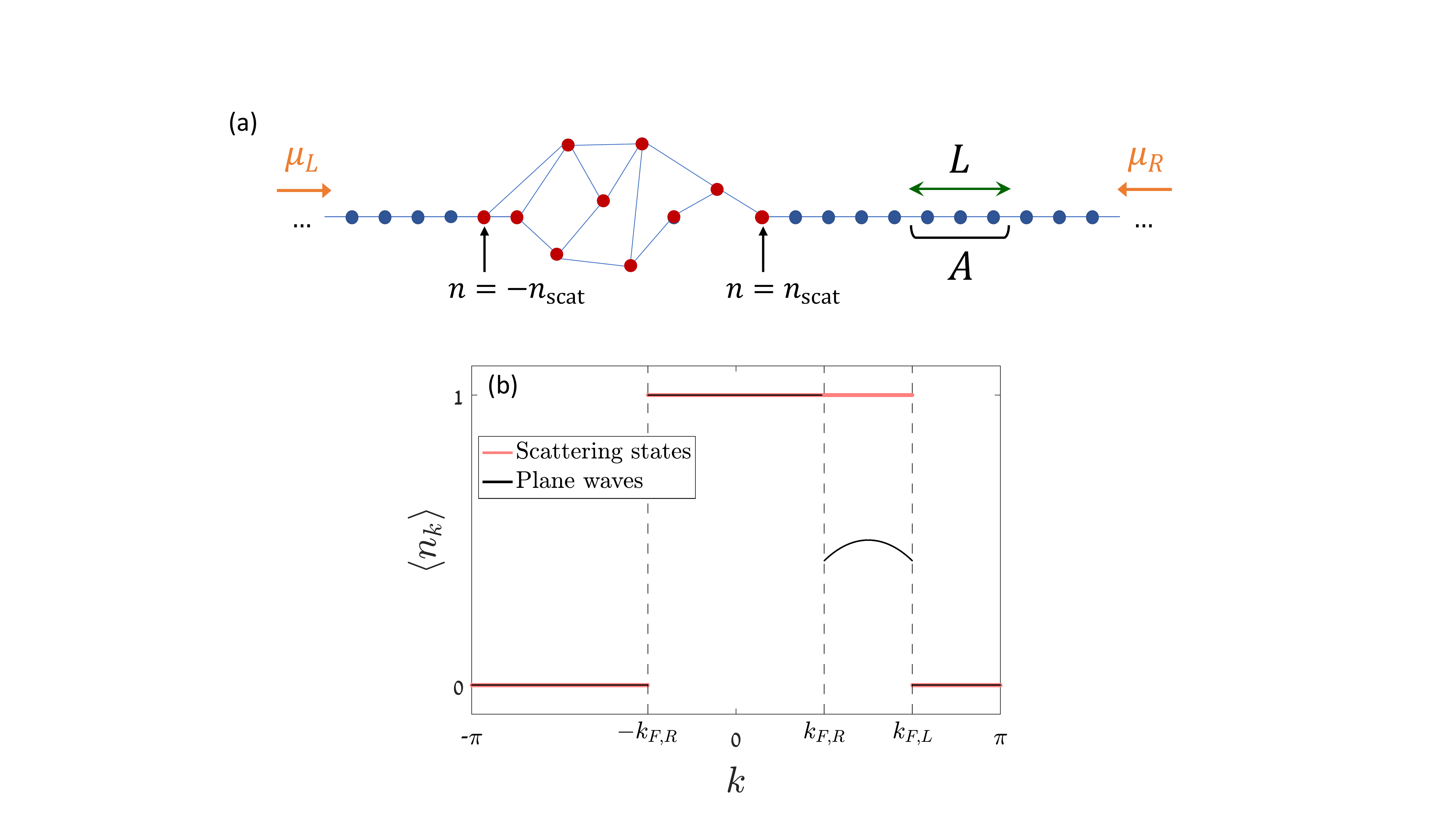}
\par\end{centering}
\caption{\label{fig:Schematic-single}(a) Schematic illustration of the general
lattice model under consideration in Secs.~\ref{sec: The Physical Model}--\ref{sec:Analytical-Asymptotics},
with the single-particle Hamiltonian given in Eq.~(\ref{eq: General Hamiltonian}).
Sites marked in blue belong to the unperturbed parts of the tight-binding
chain, while sites marked in red belong to the scattering region,
which terminates at the sites $n=\pm n_{{\rm scat}}$. $\mu_{L}$
($\mu_{R}$) designates the chemical potential for particles incoming
from the left (right). $A$ denotes the subsystem of $L$ contiguous
sites with respect to which the calculations of bipartite entanglement
in Sec.~\ref{sec:Analytical-Asymptotics} are performed. (b) Schematic
plot of the distributions as functions of lattice momentum, for the
scattering states (defined in Eqs.~(\ref{eq: Left scattering state})--(\ref{eq: Right scattering state}))
and for plane waves (in the region $n>n_{{\rm scat}}$, to the right
of the scattering region). The case shown is $k_{F,L}>k_{F,R}$.}
\end{figure}

The general model with which this paper is concerned is that of a
long homogeneous one-dimensional fermionic tight-binding chain, which
contains a finite inhomogeneous region near its middle that induces
scattering, as schematically depicted in Fig.~\ref{fig:Schematic-single}(a).
The single-particle Hamiltonian of such a system can be written as
\begin{equation}
{\cal H}=-t\sum_{n=n_{{\rm scat}}}^{N/2-1}\left(|n\rangle\langle n+1|+|-n\rangle\langle-n-1|+{\rm h.c.}\right)+{\cal H}_{{\rm scat}},\label{eq: General Hamiltonian}
\end{equation}
where $t>0$ is the hopping amplitude, $N+1$ is the number of chain
sites ($N$ is assumed to be even), and ${\cal H}_{{\rm scat}}$ is
the term that will give rise to scattering. ${\cal H}_{{\rm scat}}$
should involve only a few chain sites in the vicinity of $n=0$, and
possibly also hopping terms to additional side-attached sites in this
small region. $n_{{\rm scat}}\ge0$ is an integer such that states
$|n\rangle$ with $\left|n\right|>n_{{\rm scat}}$ do not appear in
the matrix representation of ${\cal H}_{{\rm scat}}$. Subsystem $A$,
for which we will estimate the entanglement measures, includes $L$
contiguous sites ($1\ll L\ll N$) located to one side of the scattering
region, and far away from it such that every site $n$ in $A$ obeys
$\left|n\right|-n_{{\rm scat}}\gg L$. For the sake of simplifying
the notations we shall focus on the case where the subsystem is to
the right of the scattering region, though this is of course an arbitrary
choice.

The scattering matrix \citep{merzbacher1998quantum} related to this
problem is a function of $k$, the lattice momentum,
\begin{equation}
S\left(k\right)=\left(\begin{array}{cc}
r_{L}\left(k\right) & t_{R}\left(k\right)\\
t_{L}\left(k\right) & r_{R}\left(k\right)
\end{array}\right).
\end{equation}
$S$ is unitary, and in particular $\left|r_{L}\left(k\right)\right|^{2}+\left|t_{R}\left(k\right)\right|^{2}=\left|r_{R}\left(k\right)\right|^{2}+\left|t_{L}\left(k\right)\right|^{2}=1$.
For the scattering amplitudes we use the convention that attributes
a momentum $k>0$ to a state incoming from the left and a momentum
$-k<0$ to a state incoming from the right, so that $S\left(k\right)$
is defined for $k>0$. Scattering states constitute the single-particle
energy eigenstates for energies $\left|E\right|<2t$. We assume that
the scattering potential does not support a half-bound state, i.e.~a
non-normalizable solution with energy $E=2t$ or $E=-2t$, as is the
generic case~\citep{doi:10.1063/1.525968,doi:10.1063/1.526014,Poliatzky1993}.
This condition entails that $t_{R}\left(k\right),t_{L}\left(k\right)\rightarrow0$
as $k\rightarrow0,\pi$~\citep{doi:10.1063/1.530481}. The addition
of ${\cal H}_{{\rm scat}}$ may, however, create bound states in the
single-particle energy spectrum, with energies $\left|E\right|>2t$.

If we denote by $|\psi_{k}^{\left(L\right)}\rangle$ the scattering
state related to a wave incoming from the left with momentum $k>0$,
its form outside the scattering region will be
\begin{equation}
\langle n|\psi_{k}^{\left(L\right)}\rangle=\frac{1}{\sqrt{N}}\begin{cases}
e^{ikn}+r_{L}\left(k\right)e^{-ikn} & n<-n_{{\rm scat}},\\
t_{L}\left(k\right)e^{ikn} & n>n_{{\rm scat}},
\end{cases}\label{eq: Left scattering state}
\end{equation}
while for a wave incoming from the right with momentum $-k<0$, we
will have
\begin{equation}
\langle n|\psi_{k}^{\left(R\right)}\rangle=\frac{1}{\sqrt{N}}\begin{cases}
t_{R}\left(k\right)e^{-ikn} & n<-n_{{\rm scat}},\\
e^{-ikn}+r_{R}\left(k\right)e^{ikn} & n>n_{{\rm scat}}.
\end{cases}\label{eq: Right scattering state}
\end{equation}
In the many-particle picture, the fermionic creation operator for
a site located to the right of the scattering region, $n>n_{{\rm scat}}$,
can thus be expanded as
\begin{equation}
a_{n}^{\dagger}=\frac{1}{\sqrt{N}}\sum_{k>0}\left(e^{ikn}+r_{R}\left(k\right)^{*}e^{-ikn}\right)a_{k,R}^{\dagger}+\frac{1}{\sqrt{N}}\sum_{k>0}t_{L}\left(k\right)^{*}e^{-ikn}a_{k,L}^{\dagger},
\end{equation}
where $a_{k,R}^{\dagger}$ creates the scattering state $|\psi_{k}^{\left(R\right)}\rangle$,
and $a_{k,L}^{\dagger}$ creates the scattering state $|\psi_{k}^{\left(L\right)}\rangle$.
If ${\cal H}_{{\rm scat}}$ indeed supports bound states, such states
would appear as well in the superposition of energy eigenstates that
defines $|n\rangle$. We have nevertheless ignored these states in
our writing $a_{n}^{\dagger}$ in terms of creation operators of energy
eigenstates, due to the localized nature of such bound states, which
makes their contribution to the entanglement exponentially small in
the distance of subsystem $A$ from the scattering region.

Subject to an external constant bias voltage, such a system would
arrive at a current-carrying steady state\footnote{A true steady state is reached only in the case of an infinite chain,
$N\to\infty$, which is the limit examined here. One may consider
a scenario where the system is prepared using a quench~\citep{doi:10.1063/1.3149497},
i.e.~by connecting at a specific point in time two separate leads
with different chemical potentials via the scattering region. Once
enough time has passed so that excitations that crossed from one lead
to the other have traversed the finite subsystem, the particle fluxes
incoming into and outgoing out of the subsystem become balanced, and
the entanglement properties of the subsystem relax to time-independent
values~\citep{HOOGEVEEN201578}.} where the Fermi momentum for waves incoming from the left, $k_{F,L}$,
differs from the Fermi momentum for waves incoming from the right,
$k_{F,R}$. At zero temperature this steady state is described by
a pure many-body state, 
\begin{equation}
|\Omega\rangle=\left(\prod_{0<k<k_{F,R}}a_{k,R}^{\dagger}\right)\left(\prod_{0<k<k_{F,L}}a_{k,L}^{\dagger}\right)|0\rangle,
\end{equation}
where $|0\rangle$ is the vacuum state. Taking the limit $N\to\infty$,
we replace sums over $k$ with appropriate integrals. The correlation
between two sites $m,n>n_{{\rm scat}}$ for this steady state is then
given by
\begin{equation}
C_{mn}\equiv\langle a_{m}^{\dagger}a_{n}\rangle=\frac{1}{2\pi}\underset{-\pi}{\overset{\pi}{\int}}e^{-i\left(m-n\right)k}\tau\left(k\right)dk+\frac{1}{2\pi}\underset{-\pi}{\overset{\pi}{\int}}e^{-i\left(m+n\right)k}h\left(k\right)dk,\label{eq: Correlation matrix full expression}
\end{equation}
where 
\begin{equation}
h\left(k\right)=\begin{cases}
r_{R}\left(-k\right) & -k_{F,R}<k<0,\\
r_{R}\left(k\right)^{*} & 0<k<k_{F,R},\\
0 & {\rm otherwise},
\end{cases}
\end{equation}
and, in the case where $k_{F,R}<k_{F,L}$,
\begin{equation}
\tau\left(k\right)=\begin{cases}
1 & -k_{F,R}<k<k_{F,R},\\
\left|t_{L}\left(k\right)\right|^{2} & k_{F,R}<k<k_{F,L},\\
0 & {\rm otherwise}.
\end{cases}
\end{equation}
If instead $k_{F,R}>k_{F,L}$, we obtain
\begin{equation}
\tau\left(k\right)=\begin{cases}
1 & -k_{F,R}<k<k_{F,L},\\
\left|r_{R}\left(k\right)\right|^{2} & k_{F,L}<k<k_{F,R},\\
0 & {\rm otherwise}.
\end{cases}
\end{equation}
In the case where $m,n<-n_{{\rm scat}}$, the result for $C_{mn}$
is similar up to the replacements $R\leftrightarrow L$, $\tau\left(k\right)\longrightarrow\tau\left(-k\right)$
and $h\left(k\right)\longrightarrow h\left(-k\right)$. As mentioned,
subsystem $A$ is assumed to be located to the right of the scattering
region, such that $n>n_{{\rm scat}}$ for every site in $A$. For
further convenience, we denote from now on $k_{-}=\min\left\{ k_{F,R},k_{F,L}\right\} $
and $k_{+}=\max\left\{ k_{F,R},k_{F,L}\right\} $.

For each momentum $k$, the distribution $\left\langle n_{k}\right\rangle $
of the scattering states equals either $0$ or $1$ by the definition
of the steady state. In contrast, the symbol $\tau\left(k\right)$
may be interpreted as the steady-state distribution of the unperturbed
plane waves in the region $n>n_{{\rm scat}}$; a plane wave state
within the bias voltage window $k_{-}<k<k_{+}$ will generically be
occupied with a fractional distribution factor $0<\left\langle n_{k}\right\rangle <1$.
This distinction between the distributions of the scattering states
and the plane waves is illustrated in Fig.~\ref{fig:Schematic-single}(b).
As we shall see in Subsec.~\ref{subsec:Unresolved-entanglement-analytical},
the fractional occupation of the plane waves within the voltage window
is the source of the extensive scaling of the subsystem entanglement.

We may conclude that the correlation matrix $C$ is a sum of a Toeplitz
matrix, depending on the index difference $m-n$ (the integral in
Eq.~(\ref{eq: Correlation matrix full expression}) containing $\tau\left(k\right)$),
and a Hankel matrix, depending on the sum of indexes, $m+n$ (the
integral in Eq.~(\ref{eq: Correlation matrix full expression}) containing
$h\left(k\right)$). The Hankel term is negligible for $m,n\gg n_{{\rm scat}}$
by virtue of the Riemann-Lebesgue lemma, and its decay is generically
algebraic, $\frac{1}{2\pi}\underset{-\pi}{\overset{\pi}{\int}}e^{-i\left(m+n\right)k}h\left(k\right)dk={\cal O}\left(\frac{1}{m+n}\right)$~\citep{Bender_1999}.
We can thus write
\begin{equation}
C_{mn}\approx\frac{1}{2\pi}\underset{-\pi}{\overset{\pi}{\int}}e^{-i\left(m-n\right)k}\tau\left(k\right)dk,\,\,\,\,\,\text{for }m,n\gg n_{{\rm scat}}.\label{eq: Correlation matrix approx}
\end{equation}

As we elaborate in Sec.~\ref{sec:Analytical-Asymptotics}, all analytical
calculations included in this paper rely on the approximation in Eq.~(\ref{eq: Correlation matrix approx})
of the two-site correlation matrix $C$. The use of this approximation
is justified if the distance between subsystem $A$ and the scattering
region is much larger than the length of $A$, as we further discuss
in Subsec.~\ref{subsec:Accuracy-of-analytics}. There we numerically
demonstrate the algebraic decay of the contribution of the Hankel
term (neglected within the approximation in Eq.~(\ref{eq: Correlation matrix approx}))
to the R\'enyi moments, and also find that this contribution exhibits
Friedel oscillations~\citep{Friedel1958,coleman_2015}.

\section{Analytical asymptotics of the entanglement\label{sec:Analytical-Asymptotics}}

In this section we present an analytical calculation of the R\'enyi
moments and of the vNEE for subsystem $A$ with respect to its complement.
Since in our model the total number of fermions in the lattice is
conserved, the R\'enyi moments and the vNEE may also be resolved
with respect to $Q_{A}=\sum_{m\in A}a_{m}^{\dagger}a_{m}$, the charge
in subsystem $A$. The results we pursue are more conveniently calculated
if we start by defining the following generating function~\citep{PhysRevLett.120.200602,PhysRevB.98.041106}:
\begin{equation}
Z_{n}\left(\alpha\right)={\rm Tr}\left[\rho_{A}^{n}e^{i\alpha Q_{A}}\right],\label{eq: Generating function definition}
\end{equation}
where $-\pi<\alpha<\pi$. The calculation of $Z_{n}\left(\alpha\right)$
actually encompasses all resolved and unresolved entropies and moments.
R\'enyi moments are given by $Z_{n}=Z_{n}\left(\alpha=0\right)$,
from which the vNEE can be extracted through Eq.~(\ref{eq: vNEE relation to Renyi moments}).
Furthermore, the charge-resolved R\'enyi moment is simply a Fourier
decomposition of the corresponding generating function~\citep{PhysRevLett.120.200602},
\begin{equation}
Z_{n}\left(Q_{A}\right)=\underset{-\pi}{\overset{\pi}{\int}}\frac{d\alpha}{2\pi}Z_{n}\left(\alpha\right)e^{-i\alpha Q_{A}}.\label{eq: Resolved moment from generating function}
\end{equation}
In particular, $Z_{1}\left(\alpha\right)$ is the characteristic function
of the charge distribution in $A$. The charge-resolved vNEE may be
subsequently derived through
\begin{equation}
{\cal S}\left(Q_{A}\right)=-\lim_{n\to1}\partial_{n}Z_{n}\left(Q_{A}\right).\label{eq: Resolved vNEE from resolved moment}
\end{equation}

\begin{figure}
\begin{centering}
\includegraphics[viewport=50bp 120bp 1240bp 620bp,clip,scale=0.3]{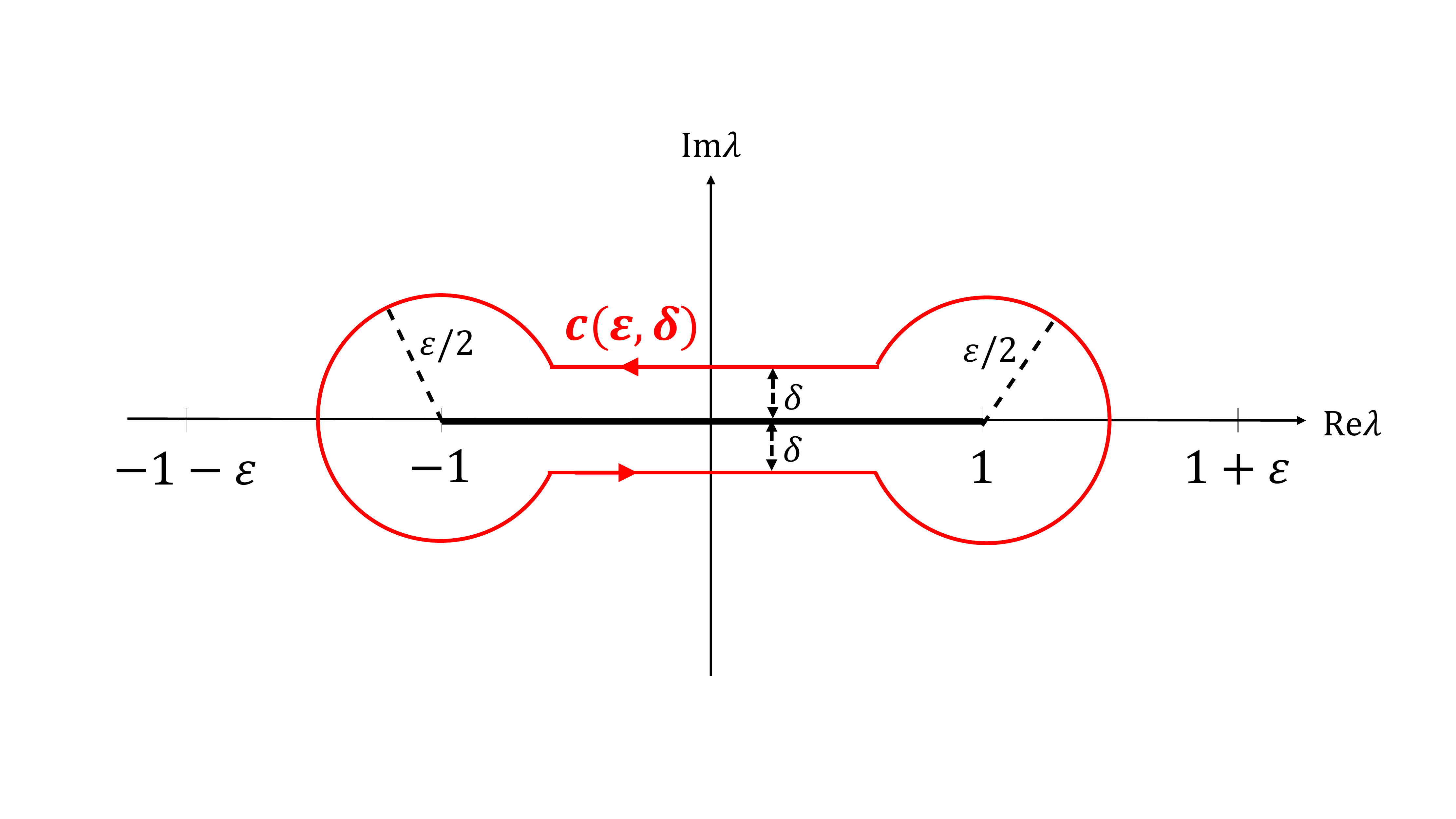}
\par\end{centering}
\caption{\label{fig: The integration contour}The integration contour used
in Eq.~(\ref{eq: Contour integral}).}
\end{figure}

$Z_{n}\left(\alpha\right)$ can be written in terms of the eigenvalues
$\left\{ \nu_{l}\right\} $ of $2C_{A}-I_{L}$, where $I_{L}$ is
the identity matrix of size $L$, and $C_{A}$ is the two-site correlation
matrix (defined as $C$ in Eq.~(\ref{eq: Correlation matrix full expression}))
restricted to subsystem $A$. More concretely, we may write~\citep{PhysRevLett.120.200602}
\begin{equation}
Z_{n}\left(\alpha\right)=\prod_{l=1}^{L}\left[\left(\frac{1+\nu_{l}}{2}\right)^{n}e^{i\alpha}+\left(\frac{1-\nu_{l}}{2}\right)^{n}\right],\label{eq: Generating function exact}
\end{equation}
which enables us to reformulate the calculation as a problem of contour
integration in the complex plane~\citep{Jin2004}:
\begin{equation}
\ln Z_{n}\left(\alpha\right)=\lim_{\varepsilon,\delta\rightarrow0^{+}}\frac{1}{2\pi i}\underset{c\left(\varepsilon,\delta\right)}{\int}e_{n}^{\left(\alpha\right)}\left(1+\varepsilon,\lambda\right)\frac{d}{d\lambda}\ln D_{L}\left(\lambda\right)d\lambda,\label{eq: Contour integral}
\end{equation}
where $D_{L}\left(\lambda\right)=\det\left(\left(\lambda+1\right)I_{L}-2C_{A}\right)$
and $e_{n}^{\left(\alpha\right)}\left(x,\nu\right)=\ln\left[\left(\frac{x+\nu}{2}\right)^{n}e^{i\alpha}+\left(\frac{x-\nu}{2}\right)^{n}\right]$.
The contour $c\left(\varepsilon,\delta\right)$ is defined such that
it encloses the segment $\left[-1,1\right]$ of the real line (on
which all the eigenvalues $\left\{ \nu_{l}\right\} $ are located)
while avoiding the singularities of $e_{n}^{\left(\alpha\right)}\left(1+\varepsilon,\lambda\right)$,
as is depicted in Fig.~\ref{fig: The integration contour}.

\subsection{Leading asymptotics of the generating function\label{subsec: Leading analytical asymptotics}}

The immediate consequence of Eq.~(\ref{eq: Correlation matrix approx})
is that $D_{L}\left(\lambda\right)$ may be approximated as a Toeplitz
determinant, meaning that $D_{L}\left(\lambda\right)=\det T_{L}\left(\lambda\right)$
where $T_{L}\left(\lambda\right)$ is a Toeplitz matrix. In particular,
$\left(T_{L}\right)_{mn}=\phi_{m-n}$ where
\begin{equation}
\phi_{l}=\frac{1}{2\pi}\underset{-\pi}{\overset{\pi}{\int}}\phi\left(k\right)e^{-ilk}dk\,\,;\,\,\phi\left(k\right)=\begin{cases}
\lambda-1 & -k_{F,R}<k<k_{-},\\
\lambda-\nu\left(k\right) & k_{-}<k<k_{+},\\
\lambda+1 & {\rm otherwise}.
\end{cases}\label{eq: Toeplitz symbol}
\end{equation}
Here we have denoted
\begin{equation}
\nu\left(k\right)=\begin{cases}
\left|t_{L}\left(k\right)\right|^{2}-\left|r_{R}\left(k\right)\right|^{2} & k_{F,R}<k_{F,L},\\
\left|r_{R}\left(k\right)\right|^{2}-\left|t_{L}\left(k\right)\right|^{2} & k_{F,R}>k_{F,L},
\end{cases}
\end{equation}
a definition that may be more compactly packed into the form
\begin{equation}
\frac{1\pm\nu\left(k\right)}{2}=\left|t_{L}\left(k\right)\right|^{2}\,\,\,\,\text{for }k_{F,L}=k_{\pm}\,\,\,\,\text{(resp. }k_{F,R}\,\text{\ensuremath{\lessgtr}}\,k_{F,L}\text{)}.\label{eq: definition of nu}
\end{equation}
Importantly, $\left|\nu\left(k\right)\right|\le1$.

The symbol $\phi\left(k\right)$ as defined in Eq.~(\ref{eq: Toeplitz symbol})
cannot be cast in the Fisher-Hartwig form~\citep{10.2307/23030524},
contrary to what is required by the well-known (and proven) formulae
and theorems of which we are aware~\citep{doi:10.1002/cpa.21467,10.2307/23030524,97a1db6c59cb4ee78c1c54c1c5610397}
concerning the asymptotics of Toeplitz determinants. A generalized
asymptotic formula for the determinant of a Toeplitz matrix generated
by a piecewise-continuous symbol was conjectured in Refs.~\citep{PhysRevA.92.042334,PhysRevA.97.062301}.
According to this formula, for $L\gg1$ we have
\begin{align}
\ln D_{L}\left(\lambda\right) & =\frac{L}{2\pi}\underset{-\pi}{\overset{\pi}{\int}}dk\ln\phi\left(k\right)\nonumber \\
 & +\frac{\ln L}{4\pi^{2}}\left[\left(\ln\frac{\lambda-1}{\lambda-\nu\left(k_{-}\right)}\right)^{2}+\left(\ln\frac{\lambda-\nu\left(k_{+}\right)}{\lambda+1}\right)^{2}+\left(\ln\frac{\lambda+1}{\lambda-1}\right)^{2}\right]+\ldots,\label{eq: Conjectured det asymptotics}
\end{align}
where the ellipses represent terms of lower order in $L$.

Plugging the asymptotic form in Eq.~(\ref{eq: Conjectured det asymptotics})
into the integral expression in Eq.~(\ref{eq: Contour integral}),
we obtain
\begin{align}
\ln Z_{n}\left(\alpha\right) & \sim\frac{L}{2\pi}\left[i\alpha\left(k_{-}+k_{F,R}\right)+\underset{k_{-}}{\overset{k_{+}}{\int}}e_{n}^{\left(\alpha\right)}\left(1,\nu\left(k\right)\right)dk\right]\nonumber \\
 & +\ln L\left[Q_{n}\left(\nu\left(k_{-}\right),\alpha\right)+Q_{n}\left(-\nu\left(k_{+}\right),-\alpha\right)+\frac{1}{12}\left(\frac{1}{n}-n\right)-\frac{\alpha^{2}}{4\pi^{2}n}\right]\nonumber \\
 & \equiv{\cal I}_{{\rm lin}}\left(n,\alpha\right)L+{\cal I}_{{\rm log}}\left(n,\alpha\right)\ln L,\label{eq: Conjectured entropy asymptotics}
\end{align}
where we have defined
\begin{equation}
Q_{n}\left(\nu,\alpha\right)=\frac{1}{2\pi^{2}}\underset{\nu}{\overset{1}{\int}}\ln\left|\frac{x-1}{x-\nu}\right|\frac{d}{dx}e_{n}^{\left(\alpha\right)}\left(1,x\right)dx.\label{eq: Discontinuity kernel notation}
\end{equation}
The term ${\cal I}_{{\rm lin}}\left(n,\alpha\right)$ is derived in
a straightforward manner, while a detailed derivation of the term
${\cal I}_{{\rm log}}\left(n,\alpha\right)$ appears in Appendix~\ref{subsec: Appendix Logarithmic Term}.

The linear term in $L$ appearing in Eq.~(\ref{eq: Conjectured entropy asymptotics})
counts the filled momentum (plane wave) states. The states with $-k_{F,R}<k<k_{-}$
are all filled with probability $1$, while if $k_{F,R}<k_{F,L}$,
the states with $k_{-}<k<k_{+}$ are filled with probability $\left|t_{L}\left(k\right)\right|^{2}$
and empty with probability $1-\left|t_{L}\left(k\right)\right|^{2}$
(or vice versa in the case where $k_{F,L}<k_{F,R}$). This distribution
is schematically presented in Fig.~\ref{fig:Schematic-single}(b).
Since any interval $\delta k$ includes $\left(L/2\pi\right)\delta k$
states, $\exp\left({\cal I}_{{\rm lin}}\left(n,\alpha\right)L\right)$
is simply the product of the moments that arise from the individual
filled states. The term $\exp\left({\cal I}_{{\rm lin}}\left(n,\alpha\right)L\right)$
can thus be interpreted as a generalization of the generating function
for the full counting statistics in the case of a transmission factor
that is constant in $k$, cf. Eq.~(20) in Ref.~\citep{doi:10.1063/1.3149497}
(the Levitov-Lesovik formula).

The logarithmic term in $L$ appearing in Eq.~(\ref{eq: Conjectured entropy asymptotics})
is a result of the Fermi discontinuities at $k=k_{\pm}$ and $k=-k_{F,R}$,
featured in Fig.~\ref{fig:Schematic-single}(b). In particular, since
$Q_{n}\left(1,\alpha\right)=0$ by definition, the contribution from
$k=k_{\pm}$ vanishes when $\nu\left(k_{\pm}\right)\rightarrow\mp1$
(respectively), in accordance with the disappearance of the respective
jump discontinuity of the symbol $\phi\left(k\right)$. That is also
the case when the bias voltage is larger than the bandwidth, such
that $k_{F,L}=0$ and $k_{F,R}=\pi$ or vice versa: since $\left|r_{R}\left(k\right)\right|\rightarrow1$
as $k\rightarrow0,\pi$, the symbol $\phi\left(k\right)$ is continuous
at any value of $k$, and thus the logarithmic term vanishes from
Eq.~(\ref{eq: Conjectured entropy asymptotics}).

\subsection{Subleading asymptotics of the generating function\label{subsec:Subleading-asymptotics-of}}

In order to incorporate further subleading terms in the analytical
asymptotics of $\ln Z_{n}\left(\alpha\right)$, we now approximate
the symbol $\phi\left(k\right)$ in Eq.~(\ref{eq: Toeplitz symbol})
to be piecewise-constant, such that it will fit the Fisher-Hartwig
form~\citep{10.2307/23030524}. The underlying assumption is that
under a small bias such that $\Delta k\equiv\left|k_{F,L}-k_{F,R}\right|\ll1$
it would be permissible to use the approximation $\nu\left(k\right)\approx\nu\left(k_{0}\right)$
for $k_{-}<k<k_{+}$, where $k_{0}\equiv\frac{1}{2}\left(k_{F,R}+k_{F,L}\right)$.
This requires the transmission and reflection factors to change slowly
near $k=k_{0}$. For this purpose we define the following approximate
symbol:
\begin{equation}
\tilde{\phi}\left(k\right)=\begin{cases}
\lambda-1 & -k_{F,R}<k<k_{-},\\
\lambda-\nu_{0} & k_{-}<k<k_{+},\\
\lambda+1 & {\rm otherwise},
\end{cases}\label{eq: Approx Toeplitz symbol}
\end{equation}
where we denoted $\nu_{0}=\nu\left(k_{0}\right)$.

Let us denote by $\tilde{D}_{L}\left(\lambda\right)$ the determinant
of the Toeplitz matrix generated by the symbol in Eq.~(\ref{eq: Approx Toeplitz symbol}).
Using the Fisher-Hartwig formula~\citep{Jin2004,10.2307/23030524}
for the asymptotics of $\tilde{D}_{L}\left(\lambda\right)$ and substituting
it into the integral expression~(\ref{eq: Contour integral}) for
the generating function will yield for $L\gg1$ an expression of the
form 
\begin{equation}
\ln Z_{n}\left(\alpha\right)\approx{\cal \tilde{I}}_{{\rm lin}}\left(n,\alpha\right)L+\tilde{{\cal I}}_{{\rm log}}\left(n,\alpha\right)\ln L+\tilde{{\cal I}}_{{\rm const}}\left(n,\alpha\right).\label{eq: Entropy FH asymptotics}
\end{equation}
Note that while the asymptotics of $\ln Z_{n}\left(\alpha\right)$
in Eq.~(\ref{eq: Conjectured entropy asymptotics}) was estimated
up to a linear term and a logarithmic term in $L$ without an approximation
of the Toeplitz symbol, here the approximation $\tilde{\phi}\left(k\right)$
yields an additional term which is independent of $L$. Compared to
the exact expressions for the linear and logarithmic terms, the error
of the terms obtained from the approximate symbol scale as
\begin{align}
{\cal I}_{{\rm lin}}\left(n,\alpha\right)-{\cal \tilde{I}}_{{\rm lin}}\left(n,\alpha\right) & \sim\left(\Delta k\right)^{3},\nonumber \\
{\cal I}_{{\rm log}}\left(n,\alpha\right)-\tilde{{\cal I}}_{{\rm log}}\left(n,\alpha\right) & \sim\left(\Delta k\right)^{2}\ln\Delta k,
\end{align}
as $\Delta k\rightarrow0$.

Relying on the Fisher-Hartwig formula, we obtain that
\begin{align}
\tilde{{\cal I}}_{{\rm const}}\left(n,\alpha\right) & =\ln\left|\frac{2\sin\frac{k_{-}+k_{F,R}}{2}\sin\frac{1}{2}\Delta k}{\sin\frac{k_{+}+k_{F,R}}{2}}\right|Q_{n}\left(\nu_{0},\alpha\right)+\ln\left|\frac{2\sin\frac{k_{+}+k_{F,R}}{2}\sin\frac{1}{2}\Delta k}{\sin\frac{k_{-}+k_{F,R}}{2}}\right|Q_{n}\left(-\nu_{0},-\alpha\right)\nonumber \\
 & +\ln\left|\frac{2\sin k_{F,R}\sin k_{0}}{\sin\frac{1}{2}\Delta k}\right|\left[\frac{1}{12}\left(\frac{1}{n}-n\right)-\frac{\alpha^{2}}{4\pi^{2}n}\right]\nonumber \\
 & +\Upsilon_{n}\left(\nu_{0},\alpha\right)+\Upsilon_{n}\left(-\nu_{0},-\alpha\right)+\Upsilon_{n}\left(-1,\alpha\right),\label{eq: FH constant term}
\end{align}
where we have defined
\begin{equation}
\Upsilon_{n}\left(\nu,\alpha\right)=\frac{1}{2\pi i}\underset{\nu}{\overset{1}{\int}}\ln\frac{\Gamma\left(\frac{1}{2}+\frac{1}{2\pi i}\ln\left(\frac{1-x}{x-\nu}\right)\right)}{\Gamma\left(\frac{1}{2}-\frac{1}{2\pi i}\ln\left(\frac{1-x}{x-\nu}\right)\right)}\frac{d}{dx}e_{n}^{\left(\alpha\right)}\left(1,x\right)dx.\label{eq: Upsilon def}
\end{equation}
We detail the calculation of $\tilde{{\cal I}}_{{\rm const}}\left(n,\alpha\right)$
in Appendix~\ref{subsec: FH constant term}. In total, the approximate
asymptotic expression for the generating function is 
\begin{equation}
\ln Z_{n}\left(\alpha\right)\approx{\cal I}_{{\rm lin}}\left(n,\alpha\right)L+{\cal I}_{{\rm log}}\left(n,\alpha\right)\ln L+\tilde{{\cal I}}_{{\rm const}}\left(n,\alpha\right).\label{eq: Generating function asymptotic expression}
\end{equation}

Note that, generically, the terms in ${\cal I}_{{\rm log}}\left(n,\alpha\right)$
and $\tilde{{\cal I}}_{{\rm const}}\left(n,\alpha\right)$ which stem
from the partial transmission effects for $k_{-}<k<k_{+}$ do not
vanish when we fix $L$ and simply take the limit $\Delta k\rightarrow0$;
we must first take $\nu\left(k\right)\rightarrow\pm1$ in the interval
$k_{-}<k<k_{+}$ in order for them to vanish. This, however, is in
compliance with the fact that the problem is examined at the limit
of large $L$, which thus constitutes the largest length scale of
the problem (other than the length scales that are assumed infinite,
i.e.~the length of the full chain and the distance of $A$ from the
scattering region). The limits $L\to\infty$ and $\Delta k\to0$ do
not commute; in other words, our expressions assume $\Delta k\gg1/L$,
and hence taking $\Delta k\to0$ in them does not eliminate the contributions
coming from the partial transmission within the voltage window $\Delta k$.

A more problematic feature of the analytical result is that for any
$n$, the function $Q_{n}\left(\nu,\alpha\right)$ (which appears
in the expressions for both ${\cal I}_{{\rm log}}\left(n,\alpha\right)$
and $\tilde{{\cal I}}_{{\rm const}}\left(n,\alpha\right)$) is singular
at $\nu=0$ and $\alpha=\pm\pi$. Moreover, it may be shown that
\begin{equation}
{\rm Re}Q_{n}\left(0,\alpha\right)\sim\frac{1}{4\pi^{2}}\ln^{2}\left(\frac{\pi\mp\alpha}{2n}\right),\,\,\,\,\,\text{as }\alpha\to\pm\pi.\label{eq: Divergence at nu=00003D0}
\end{equation}
This deems the analytical expression for $Z_{n}\left(\alpha\right)$
to be a non-integrable function of $\alpha$ over $\left[-\pi,\pi\right]$
if either $\nu\left(k_{+}\right)=0$ or $\nu\left(k_{-}\right)=0$.
Numerical results do not exhibit this kind of divergence, and so this
property of the analytical result does not capture the true behavior
of the generating function. The singularity of the function $e_{n}^{\left(\alpha\right)}\left(1,\nu\right)$
(defined right after Eq.~(\ref{eq: Contour integral})) at $\nu=0,\alpha=\pm\pi$
is what brings about this difficulty, as small shifts of $\nu\left(k_{\pm}\right)$
become crucial when either one is near $\nu=0$.

We note that our previous work~\citep{Fraenkel_2020}, which had
discussed a situation where the Toeplitz symbol may indeed be cast
in the Fisher-Hartwig form, established that corrections to the approximation
of $\ln Z_{n}\left(\alpha\right)$ using the Fisher-Hartwig formula
decay less rapidly with $L$ as $\alpha$ nears $\pm\pi$. At $\alpha=\pm\pi$
these corrections eventually become as important as the terms in Eq.~(\ref{eq: Entropy FH asymptotics}),
causing a considerable deviation from exact numerical results if one
does not include the corrections~\citep{Bonsignori_2019,Fraenkel_2020,Capizzi_2020}.
Although there is no known expression for these corrections when the
Toeplitz symbol does not fit the Fisher-Hartwig form, we expect them
to eliminate the divergence at $\alpha=\pm\pi$ observed in this case.
While the divergence of the generating function prevents us from obtaining
analytical results for charge-resolved quantities in cases where either
$\nu\left(k_{+}\right)$ or $\nu\left(k_{-}\right)$ exactly vanish,
we have found that it has little effect whenever $\nu\left(k_{\pm}\right)$
are finite, even when they are small, as demonstrated in Subsec.~\ref{subsec:Single-impurity-resolved}.

We may also estimate the deviation of the generating function for
the nonequilibrium steady state from that of the ground state in the
equilibrium case. In Refs.~\citep{Bonsignori_2019,Fraenkel_2020}
it was shown that the leading-order asymptotics of the generating
function for the ground state of a homogeneous tight-binding chain
filled up to $k=\pm k_{0}$ is given by\footnote{This is equal to the expression obtained for the nonequilibrium steady
state with a nonzero bias voltage, but in the absence of scattering,
i.e.~assuming $t_{L}\left(k\right)=1$ for all $k$.}
\begin{equation}
\ln Z_{n}^{{\rm eq}}\left(\alpha\right)\approx i\frac{k_{0}\alpha}{\pi}L+\left[\frac{1}{6}\left(\frac{1}{n}-n\right)-\frac{\alpha^{2}}{2\pi^{2}n}\right]\ln\left|2L\sin k_{0}\right|+2\Upsilon_{n}\left(-1,\alpha\right).\label{eq: Equilibrium generating function}
\end{equation}
Subtracting this from the nonequilibrium result, we obtain
\begin{align}
\ln\frac{Z_{n}\left(\alpha\right)}{Z_{n}^{{\rm eq}}\left(\alpha\right)} & \approx\frac{L}{2\pi}\left[i\alpha\left(k_{-}-k_{F,L}\right)+\underset{k_{-}}{\overset{k_{+}}{\int}}e_{n}^{\left(\alpha\right)}\left(1,\nu\left(k\right)\right)dk\right]\nonumber \\
 & +\ln L\left[Q_{n}\left(\nu\left(k_{-}\right),\alpha\right)+Q_{n}\left(-\nu\left(k_{+}\right),-\alpha\right)-\frac{1}{12}\left(\frac{1}{n}-n\right)+\frac{\alpha^{2}}{4\pi^{2}n}\right]\nonumber \\
 & +\ln\left|\frac{2\sin\frac{k_{-}+k_{F,R}}{2}\sin\frac{1}{2}\Delta k}{\sin\frac{k_{+}+k_{F,R}}{2}}\right|Q_{n}\left(\nu_{0},\alpha\right)+\ln\left|\frac{2\sin\frac{k_{+}+k_{F,R}}{2}\sin\frac{1}{2}\Delta k}{\sin\frac{k_{-}+k_{F,R}}{2}}\right|Q_{n}\left(-\nu_{0},-\alpha\right)\nonumber \\
 & -\ln\left|\frac{2\sin\frac{1}{2}\Delta k\sin k_{0}}{\sin k_{F,R}}\right|\left[\frac{1}{12}\left(\frac{1}{n}-n\right)-\frac{\alpha^{2}}{4\pi^{2}n}\right]\nonumber \\
 & +\Upsilon_{n}\left(\nu_{0},\alpha\right)+\Upsilon_{n}\left(-\nu_{0},-\alpha\right)-\Upsilon_{n}\left(-1,\alpha\right).\label{eq:genfun-nonequilibrium-deviation}
\end{align}

\subsection{Charge statistics\label{subsec:Charge-statistics}}

An important special case of the symmetry-resolved R\'enyi moments
is that of $n=1$, since $Z_{1}\left(Q_{A}\right)$ constitutes the
charge distribution in subsystem $A$, and an expansion of $\ln Z_{1}\left(\alpha\right)$
in powers of $\alpha$ gives its moments. Indeed, we may write

\begin{equation}
\ln\frac{Z_{1}\left(\alpha\right)}{Z_{1}^{{\rm eq}}\left(\alpha\right)}=i\left(\left\langle Q_{A}\right\rangle -\left\langle Q_{A}\right\rangle _{{\rm eq}}\right)\alpha-\frac{1}{2}\left[\left(\Delta Q_{A}\right)^{2}-\left(\Delta Q_{A}\right)_{{\rm eq}}^{2}\right]\alpha^{2}+{\cal O}\left(\alpha^{3}\right),\label{eq: REE alpha expansion}
\end{equation}
 where
\begin{equation}
\left\langle Q_{A}\right\rangle -\left\langle Q_{A}\right\rangle _{{\rm eq}}=-\frac{1}{2\pi}\left[\underset{k_{F,R}}{\overset{k_{F,L}}{\int}}\left|r_{R}\left(k\right)\right|^{2}dk\right]L\label{eq: Mean charge shift}
\end{equation}
is the shift in the mean charge, and
\begin{align}
\left(\Delta Q_{A}\right)^{2}-\left(\Delta Q_{A}\right)_{{\rm eq}}^{2} & =\frac{1}{2\pi}\left(\underset{k_{-}}{\overset{k_{+}}{\int}}\left|t_{L}\left(k\right)r_{R}\left(k\right)\right|^{2}dk\right)L\nonumber \\
 & -\frac{1}{2\pi^{2}}\left(1-\left|r_{R}\left(k_{F,R}\right)\right|^{4}-\left|t_{L}\left(k_{F,L}\right)\right|^{4}\right)\ln L\nonumber \\
 & +\frac{\left|r_{R}\left(k_{0}\right)\right|^{2}}{\pi^{2}}\ln\left|\frac{\sin k_{F,R}}{\sin k_{0}}\right|-\frac{\left|t_{L}\left(k_{0}\right)r_{R}\left(k_{0}\right)\right|^{2}}{\pi^{2}}\left(1+\gamma_{E}+\ln\left|2\sin\frac{\Delta k}{2}\right|\right)\label{eq: Charge variance shift}
\end{align}
is the shift in the charge variance. Here $\gamma_{E}\approx0.577$
is the Euler-Mascheroni constant~\citep{whittaker_watson_1996}.
A derivation of Eqs.~(\ref{eq: Mean charge shift}) and~(\ref{eq: Charge variance shift})
appears in Appendix~\ref{subsec: Expansion of generating function}.
Let us note that the equilibrium values of the mean and variance of
the charge in subsystem $A$ are~\citep{Fraenkel_2020} 
\begin{equation}
\left\langle Q_{A}\right\rangle _{{\rm eq}}=\frac{k_{0}}{\pi}L\,\,\,,\,\,\,\left(\Delta Q_{A}\right)_{{\rm eq}}^{2}=\frac{\ln\left|2L\sin k_{0}\right|+1+\gamma_{E}}{\pi^{2}}.
\end{equation}

Analogously, we may define a generalized quantity $\left\langle Q_{A}\right\rangle _{n}=-i\partial_{\alpha}\ln Z_{n}\left(\alpha\right)\vert_{\alpha=0}$,
designating the mean of the ``charge distribution'' whose characteristic
function is $Z_{n}\left(\alpha\right)$. By taking the derivative
of Eq.~(\ref{eq: Generating function asymptotic expression}), this
generalized mean charge is found to be
\begin{align}
\left\langle Q_{A}\right\rangle _{n} & =\left[k_{-}+k_{F,R}+\underset{k_{-}}{\overset{k_{+}}{\int}}\frac{\left(1+\nu\left(k\right)\right)^{n}}{\left(1+\nu\left(k\right)\right)^{n}+\left(1-\nu\left(k\right)\right)^{n}}dk\right]\frac{L}{2\pi}\nonumber \\
 & +\left[\underset{\nu\left(k_{-}\right)}{\overset{1}{\int}}\ln\left|\frac{x-1}{x-\nu\left(k_{-}\right)}\right|\mathfrak{g}_{n}\left(x\right)dx-\underset{-\nu\left(k_{+}\right)}{\overset{1}{\int}}\ln\left|\frac{x-1}{x+\nu\left(k_{+}\right)}\right|\mathfrak{g}_{n}\left(x\right)dx\right]\frac{\ln L}{\pi^{2}}\nonumber \\
 & +\frac{1}{\pi^{2}}\ln\left|\frac{2\sin\frac{k_{-}+k_{F,R}}{2}\sin\frac{1}{2}\Delta k}{\sin\frac{k_{+}+k_{F,R}}{2}}\right|\underset{\nu_{0}}{\overset{1}{\int}}\ln\left|\frac{x-1}{x-\nu_{0}}\right|\mathfrak{g}_{n}\left(x\right)dx\nonumber \\
 & -\frac{1}{\pi^{2}}\ln\left|\frac{2\sin\frac{k_{+}+k_{F,R}}{2}\sin\frac{1}{2}\Delta k}{\sin\frac{k_{-}+k_{F,R}}{2}}\right|\underset{-\nu_{0}}{\overset{1}{\int}}\ln\left|\frac{x-1}{x+\nu_{0}}\right|\mathfrak{g}_{n}\left(x\right)dx\nonumber \\
 & +\frac{1}{\pi i}\underset{\nu_{0}}{\overset{1}{\int}}\ln\frac{\Gamma\left(\frac{1}{2}+\frac{1}{2\pi i}\ln\left(\frac{1-x}{x-\nu_{0}}\right)\right)}{\Gamma\left(\frac{1}{2}-\frac{1}{2\pi i}\ln\left(\frac{1-x}{x-\nu_{0}}\right)\right)}\mathfrak{g}_{n}\left(x\right)dx-\frac{1}{\pi i}\underset{-\nu_{0}}{\overset{1}{\int}}\ln\frac{\Gamma\left(\frac{1}{2}+\frac{1}{2\pi i}\ln\left(\frac{1-x}{x+\nu_{0}}\right)\right)}{\Gamma\left(\frac{1}{2}-\frac{1}{2\pi i}\ln\left(\frac{1-x}{x+\nu_{0}}\right)\right)}\mathfrak{g}_{n}\left(x\right)dx,\label{eq: Generalized mean charge}
\end{align}
\sloppy where we have denoted $\mathfrak{g}_{n}\left(x\right)=\frac{n\left(1-x^{2}\right)^{n-1}}{\left[\left(1+x\right)^{n}+\left(1-x\right)^{n}\right]^{2}}$.
In similar fashion, one can obtain an analytical expression for the
corresponding variance by calculating $\left(\Delta Q_{A}\right)_{n}^{2}=-\partial_{\alpha}^{2}\ln Z_{n}\left(\alpha\right)\vert_{\alpha=0}$,
and in particular find that generically it scales linearly with $L$.

\subsection{Unresolved entanglement \label{subsec:Unresolved-entanglement-analytical}}

By setting $\alpha=0$ in the generating function from Eq.~(\ref{eq: Generating function asymptotic expression}),
we obtain analytical expressions for the unresolved entanglement measures
we wished to estimate. R\'enyi moments and entropies are directly
accessible in this manner, while the vNEE is extracted through its
relation to the R\'enyi moments, per Eq.~(\ref{eq: vNEE relation to Renyi moments}).
In order to conveniently present the resultant asymptotics for the
vNEE, we define the following functions:
\begin{align}
q\left(p\right) & =\frac{1}{8}-\frac{p}{24}-\frac{1}{2\pi^{2}}\underset{0}{\overset{1}{\int}}\frac{dx}{x}\left\{ \frac{\left(1+px\right)\ln\left(1+px\right)+\left(x+p\right)\ln\left(x+p\right)}{1+x}-p\ln p\right\} ,\nonumber \\
\upsilon\left(p\right) & =\kappa_{0}-\frac{1}{2\pi^{2}}\underset{0}{\overset{1}{\int}}\frac{dx}{x}\left\{ \frac{\left(1+px\right)\ln\left(1+px\right)+\left(x+p\right)\ln\left(x+p\right)+\left(1-p\right)x\ln x}{1+x}-p\ln p\right\} \nonumber \\
 & \times\underset{0}{\overset{\infty}{\int}}\left[\frac{\cos\left(\frac{\ln x}{2\pi}z\right)}{2\sinh\left(\frac{z}{2}\right)}-\frac{e^{-z}}{z}\right]dz,\label{eq: Auxiliary functions vNEE}
\end{align}
where we have introduced the constant $\kappa_{0}=\underset{0}{\overset{\infty}{\int}}\left[\frac{1}{z^{2}\sinh\left(\frac{z}{2}\right)}-\frac{1}{2z\sinh^{2}\left(\frac{z}{2}\right)}-\frac{e^{-z}}{12z}\right]dz\approx0.1399$.

We then have for the vNEE the following result:
\begin{equation}
{\cal S}\sim{\cal C}_{{\rm lin}}L+{\cal C}_{{\rm log}}\ln L+{\cal C}_{{\rm const}},\label{eq: Unresolved vNEE asymptotics}
\end{equation}
where ${\cal C}_{{\rm lin}}$ and ${\cal C}_{{\rm log}}$ are both
exact and are given by
\begin{equation}
{\cal C}_{{\rm lin}}=-\frac{1}{2\pi}\underset{k_{-}}{\overset{k_{+}}{\int}}\left[\left|t_{L}\left(k\right)\right|^{2}\ln\left(\left|t_{L}\left(k\right)\right|^{2}\right)+\left|r_{R}\left(k\right)\right|^{2}\ln\left(\left|r_{R}\left(k\right)\right|^{2}\right)\right]dk\label{eq: vNEE linear coefficient}
\end{equation}
and
\begin{equation}
{\cal C}_{{\rm log}}=\frac{1}{6}+q\left(\left|t_{L}\left(k_{F,R}\right)\right|^{2}\right)+q\left(\left|r_{R}\left(k_{F,L}\right)\right|^{2}\right),\label{eq: vNEE log coefficient}
\end{equation}
and ${\cal C}_{{\rm const}}$ is the approximate constant correction
(valid, as before, when $\Delta k$ is small enough),
\begin{align}
{\cal C}_{{\rm const}} & =\ln\left|\frac{2\sin k_{F,R}\sin\frac{1}{2}\Delta k}{\sin k_{0}}\right|q\left(\left|t_{L}\left(k_{0}\right)\right|^{2}\right)+\ln\left|\frac{2\sin k_{0}\sin\frac{1}{2}\Delta k}{\sin k_{F,R}}\right|q\left(\left|r_{R}\left(k_{0}\right)\right|^{2}\right)\nonumber \\
 & +\frac{1}{6}\ln\left|\frac{2\sin k_{F,R}\sin k_{0}}{\sin\frac{1}{2}\Delta k}\right|+\upsilon\left(\left|t_{L}\left(k_{0}\right)\right|^{2}\right)+\upsilon\left(\left|r_{R}\left(k_{0}\right)\right|^{2}\right)+\upsilon\left(0\right).\label{eq: vNEE const coefficient}
\end{align}
This is derived with further details in Appendix~\ref{subsec: The von-Neumman-entanglement}.
Eq.~(\ref{eq: Unresolved vNEE asymptotics}) was already highlighted
in Sec.~\ref{sec:Introduction} (where it is featured as Eq.~(\ref{eq: main scaling law}))
as a central result of this work.

The form of ${\cal C}_{{\rm lin}}$ in Eq.~(\ref{eq: vNEE linear coefficient})
is especially illuminating. The extensive term of the entanglement
entropy arises from the integration of a classical mixture entropy
with respect to the reflection and transmission probabilities. It
also highlights the two crucial ingredients that produce the linear
leading term: the nonequilibrium setting brought about by the bias
voltage, which ensures that $k_{+}\neq k_{-}$ and therefore that
the integral does not trivially vanish; and a scattering potential
that generates imperfect transmission, seeing that a unity transmission
factor will cause the integrand in Eq.~(\ref{eq: vNEE linear coefficient})
to vanish for all $k$. Both conditions must apply in order for the
leading term of ${\cal S}$ to be extensive (for related treatments
of particular time-dependent impurity setups, see Refs.~\citep{Eisler_2012,10.21468/SciPostPhys.8.3.036}).

\subsection{Charge-resolved entanglement}

An exact computation of the charge-resolved R\'enyi moments based
on the analytical asymptotics of the generating function in Eq.~(\ref{eq: Generating function asymptotic expression})
requires carrying out the integration in Eq.~(\ref{eq: Resolved moment from generating function}),
which cannot itself be performed analytically. For $L\gg1$, a useful
approximation is obtained by expanding $\ln Z_{n}\left(\alpha\right)$
in powers of $\alpha$ up to second order and replacing the integration
limits in Eq.~(\ref{eq: Resolved moment from generating function})
by $\pm\infty$. This leads to an approximate Gaussian form of the
charge-resolved $n$th R\'enyi moment:
\begin{equation}
Z_{n}\left(Q_{A}\right)\approx\frac{Z_{n}}{\sqrt{2\pi\left(\Delta Q_{A}\right)_{n}^{2}}}\exp\left[-\frac{\left(Q_{A}-\left\langle Q_{A}\right\rangle _{n}\right)^{2}}{2\left(\Delta Q_{A}\right)_{n}^{2}}\right].\label{eq: Renyi moment Gaussian}
\end{equation}
The above approximation holds since for large $L$, $\left(\Delta Q_{A}\right)_{n}^{2}$
scales linearly with $L$ (as mentioned in Subsec.~\ref{subsec:Charge-statistics}),
and consequently $Z_{n}\left(\alpha\right)$ decays rapidly away from
$\alpha=0$ (this is analogous to the central limit theorem).

The Gaussian approximation allows us to analytically examine the question
of entanglement equipartition. By plugging Eq.~(\ref{eq: Renyi moment Gaussian})
into Eq.~(\ref{eq: Definition for post measurement vNEE}), we obtain
the following expression for the vNEE after a projective charge measurement:
\begin{align}
\sigma\left(Q_{A}\right) & \approx\mathcal{S}-\frac{1}{2}\ln\left(2\pi\left(\Delta Q_{A}\right)^{2}\right)-\frac{\partial_{n}\left\langle Q_{A}\right\rangle _{n}\vert_{n=1}}{\left(\Delta Q_{A}\right)^{2}}\left(Q_{A}-\left\langle Q_{A}\right\rangle \right)\nonumber \\
 & -\frac{1}{2}\left(\frac{Q_{A}-\left\langle Q_{A}\right\rangle }{\Delta Q_{A}}\right)^{2}+\frac{\partial_{n}\left(\Delta Q_{A}\right)_{n}\vert_{n=1}}{\Delta Q_{A}}\left(1-\left(\frac{Q_{A}-\left\langle Q_{A}\right\rangle }{\Delta Q_{A}}\right)^{2}\right).\label{eq: Result for post measurement vNEE}
\end{align}
Relying on Eqs.~(\ref{eq: Charge variance shift}) and~(\ref{eq: Generalized mean charge}),
we note that both $\left(\Delta Q_{A}\right)^{2}$ and $\partial_{n}\left\langle Q_{A}\right\rangle _{n}\vert_{n=1}$
scale linearly with $L$ to leading order. In particular, the approximation
in Eq.~(\ref{eq: Result for post measurement vNEE}) is expected
to be valid for values of $Q_{A}$ obeying $Q_{A}-\left\langle Q_{A}\right\rangle ={\cal O}\left(\sqrt{L}\right)$.
Eq.~(\ref{eq: Result for post measurement vNEE}) entails that, to
leading (linear in $L$) order, entanglement is spread equally among
charge sectors with $\sigma\left(Q_{A}\right)\approx{\cal S}$, but
also that this equipartition may be broken by a term up to order ${\cal O}\left(\sqrt{L}\right)$.
To the best of our knowledge, this is the first calculation of symmetry-resolved
entanglement entropy showing a term breaking equipartition that grows
with the size of the subsystem in question~\citep{Bonsignori_2019,Fraenkel_2020,10.21468/SciPostPhys.8.3.046,PhysRevB.102.014455,Murciano_2020,Capizzi_2020,PhysRevB.103.L041104}.

The fact that the first term breaking entanglement equipartition is
odd with respect to $Q_{A}-\left\langle Q_{A}\right\rangle $ is noteworthy
as well. Previous works that explicitly calculated the first equipartition-breaking
term in different equilibrium and nonequilibrium models have always
found it to be an even function of the deviation from the mean charge~\citep{Bonsignori_2019,Fraenkel_2020,10.21468/SciPostPhys.8.3.046,PhysRevB.102.014455,Murciano_2020,Capizzi_2020,PhysRevB.103.L041104}.
For $L\gg1$ this odd term is given by
\begin{equation}
-\frac{\partial_{n}\left\langle Q_{A}\right\rangle _{n}\vert_{n=1}}{\left(\Delta Q_{A}\right)^{2}}\left(Q_{A}-\left\langle Q_{A}\right\rangle \right)\approx\frac{\underset{k_{-}}{\overset{k_{+}}{\int}}\left(1-\nu\left(k\right)^{2}\right)\ln\frac{1-\nu\left(k\right)}{1+\nu\left(k\right)}dk}{\underset{k_{-}}{\overset{k_{+}}{\int}}\left(1-\nu\left(k\right)^{2}\right)dk}\left(Q_{A}-\left\langle Q_{A}\right\rangle \right),\label{eq: Equipartition breaking slope}
\end{equation}
which suggests that, for a small nonzero bias voltage, the post-measurement
vNEEs of two charge sectors $Q_{A}=q_{1}$ and $Q_{A}=q_{2}$ approximately
differ by
\begin{equation}
\sigma\left(q_{1}\right)-\sigma\left(q_{2}\right)\approx\ln\left(\frac{1-\nu_{0}}{1+\nu_{0}}\right)\left(q_{1}-q_{2}\right).
\end{equation}

In the cases of either zero bias voltage ($\Delta k=0$) or perfect
transmission or reflection ($t_{L}\left(k\right)=1$ or $t_{L}\left(k\right)=0$,
respectively, for all $k$), we have $\partial_{n}\left\langle Q_{A}\right\rangle _{n}=0$,
so the equipartition-breaking term that displays both these novel
features vanishes. We stress that these features are unique even with
respect to the case studied in Ref.~\citep{PhysRevB.103.L041104},
where entanglement equipartition is examined for a nonequilibrium
steady state created following a global quench. The main distinction
between the steady state in Ref.~\citep{PhysRevB.103.L041104} and
the steady state we investigated here is that the latter is a state
with a net current that is partially transmitted by the scatterer.

\section{The single impurity model\label{sec: The-Single-Impurity}}

We consider a concrete example of the general model discussed above,
by setting an on-site energy cost for the middle site of the chain.
The single-particle Hamiltonian in Eq.~(\ref{eq: General Hamiltonian})
becomes
\begin{equation}
{\cal H}=-t\sum_{n=-\infty}^{\infty}\left(|n\rangle\langle n+1|+|n+1\rangle\langle n|\right)+\epsilon_{0}|0\rangle\langle0|,
\end{equation}
where $\epsilon_{0}\in\mathbb{R}$. The scattering states which constitute
solutions for the single-particle problem provide the following transmission
and reflection coefficients:
\begin{equation}
\left|t_{R,L}\left(k\right)\right|^{2}=\frac{\sin^{2}k}{\sin^{2}k+\left(\epsilon_{0}/2t\right)^{2}}\,\,\,\,\,,\,\,\,\,\,\left|r_{R,L}\left(k\right)\right|^{2}=\frac{\left(\epsilon_{0}/2t\right)^{2}}{\sin^{2}k+\left(\epsilon_{0}/2t\right)^{2}}.
\end{equation}
There is also a bound state with energy $E>2t$ ($E<-2t$) for $\epsilon_{0}>0$
($\epsilon_{0}<0$); however, as noted above, its contribution is
negligible in the limit considered. The generating function $Z_{n}\left(\alpha\right)$
thus depends on the parameters $k_{\pm}$ and $\epsilon_{0}/t$, along
with the more explicit dependence on $\alpha$ and $n$. Appendix~\ref{sec:Appendix:-Additional-plots}
illustrates how the coefficients that define the analytical asymptotic
expression for $\ln Z_{n}\left(\alpha\right)$ in Eq.~(\ref{eq: Generating function asymptotic expression})
vary with these parameters.

In this section we first focus on measures of entanglement extracted
from the generating function for the single impurity model, with a
comparison of our analytical results to numerics (Subsecs.~\ref{subsec:Single-impurity-unresolved}
and~\ref{subsec:Single-impurity-resolved}). We then use numerics
for this model to discuss more generally the accuracy of the calculation
of the generating function (Subsec.~\ref{subsec:Accuracy-of-analytics}).

\begin{figure}[t]
\begin{centering}
\includegraphics[viewport=120bp 0bp 1240bp 604bp,clip,scale=0.38]{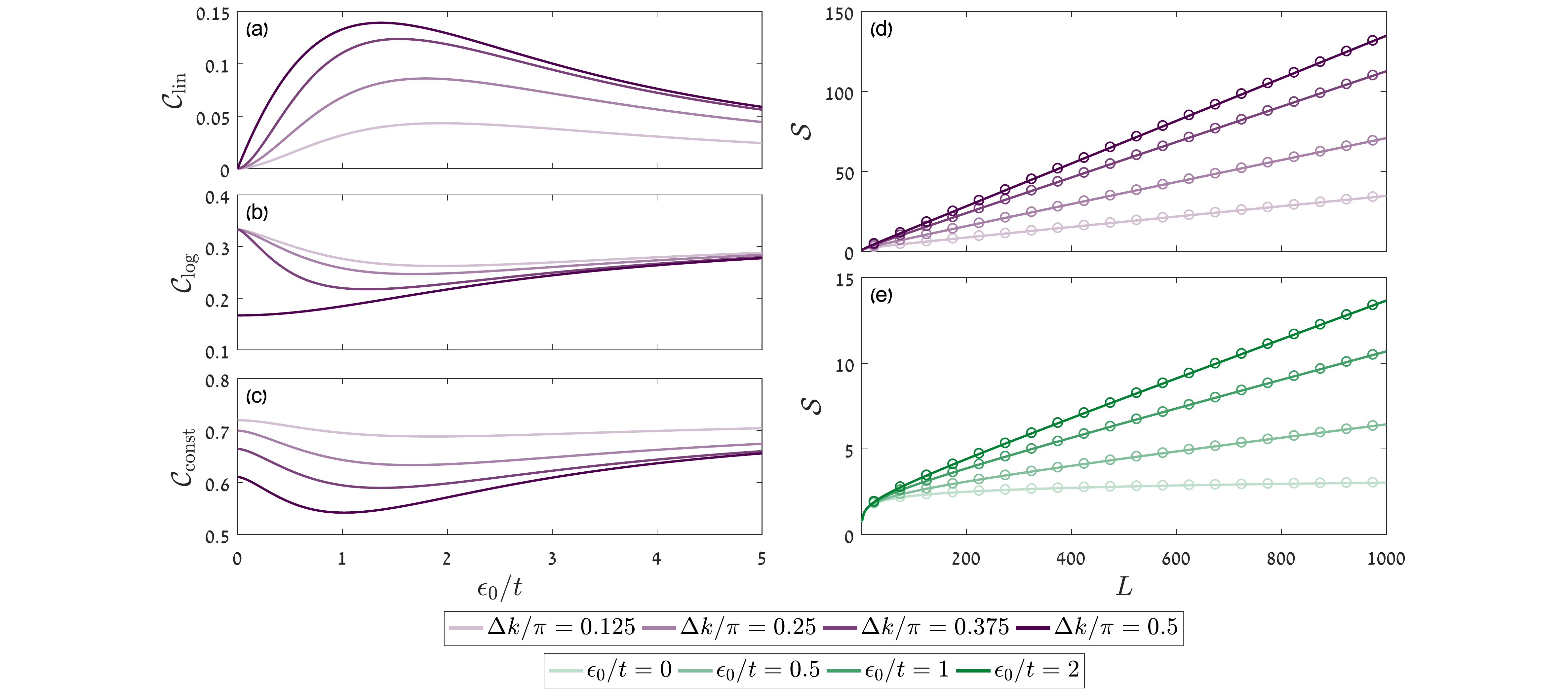}
\par\end{centering}
\caption{\label{fig: Single impurity unresolved vNEE}The single impurity model:
Unresolved vNEE. (a)--(c) Asymptotic scaling coefficients, as defined
in Eqs.~(\ref{eq: vNEE linear coefficient})--(\ref{eq: vNEE const coefficient}),
as functions of $\epsilon_{0}/t$ for various fixed values of $\Delta k=k_{F,L}-k_{F,R}$,
with $k_{F,R}=\pi/2$. (d)--(e) The vNEE ${\cal S}$ as a function
of the length $L$ of subsystem $A$, calculated analytically (lines)
using Eq.~(\ref{eq: Unresolved vNEE asymptotics}), and numerically
(circles) using Eqs.~(\ref{eq: vNEE relation to Renyi moments}),
(\ref{eq: Correlation matrix approx}) and~(\ref{eq: Generating function exact}).
In (d) we show results for various fixed values of $\Delta k=k_{F,L}-k_{F,R}$,
with $\epsilon_{0}=t$ and $k_{F,R}=\pi/2$; in (e) we show results
for various fixed values of $\epsilon_{0}/t$, with $k_{F,R}=\pi/2$
and $k_{F,L}-k_{F,R}=0.1$.}
\end{figure}

\subsection{Unresolved von-Neumann entanglement entropy\label{subsec:Single-impurity-unresolved}}

As already stressed, the entanglement between subsystem $A$ and its
complement is most rigorously quantified by the vNEE. In Fig.~\ref{fig: Single impurity unresolved vNEE}
we plot the dependence on the model parameters of the asymptotic scaling
coefficients of the vNEE from Eq.~(\ref{eq: Unresolved vNEE asymptotics}).
It can be seen that the coefficients are nonmonotonic in $\epsilon_{0}/t$.
Notably, the integral form of ${\cal C}_{{\rm lin}}$ in Eq.~(\ref{eq: vNEE linear coefficient})
suggests that the largest contribution to the leading extensive term
of ${\cal S}$ comes form momentum states where $\left|t_{L}\left(k\right)\right|^{2}\approx\left|r_{R}\left(k\right)\right|^{2}$;
this, in turn, implies that ${\cal C}_{{\rm lin}}$ should peak at
a value of $\epsilon_{0}/t$ such that $\left|t_{L}\left(k_{0}\right)\right|^{2}\approx\left|r_{R}\left(k_{0}\right)\right|^{2}$,
as is evident in Fig.~\ref{fig: Single impurity unresolved vNEE}(a).

Another noteworthy detail is that Fig.~\ref{fig: Single impurity unresolved vNEE}(b)
exemplifies the non-continuous nature of the asymptotic scaling coefficients
that depend on the Fermi discontinuities. Indeed, in a homogeneous
chain ($\epsilon_{0}=0$) we have ${\cal C}_{{\rm log}}=1/3$ for
any $0<k_{0}<\pi$, but Fig.~\ref{fig: Single impurity unresolved vNEE}(b)
shows that by fixing $k_{F,L}=\pi$ first and taking the limit $\epsilon_{0}/t\to0$
later, we obtain ${\cal C}_{{\rm log}}\to1/6$. This is because for
any $\epsilon_{0}\neq0$ there is no Fermi discontinuity at $k_{F,L}=\pi$,
but one is created at $\epsilon_{0}=0$. One must therefore be careful
when taking limits that either create or destroy jumps in the distribution.

To corroborate these analytical results, the vNEE was also extracted
through Eq.~(\ref{eq: vNEE relation to Renyi moments}) from a numerical
calculation of the R\'enyi moments $Z_{n}$. The latter is performed
using the exact expression for the generating function in Eq.~(\ref{eq: Generating function exact}),
where the restricted correlation matrix $C_{A}$ is approximated according
to Eq.~(\ref{eq: Correlation matrix approx}), so that effects of
a finite distance between subsystem $A$ and the impurity site are
neglected. Figs.~\ref{fig: Single impurity unresolved vNEE}(d)--(e)
feature a comparison of the numerical result for the vNEE with our
analytical calculation, confirming good agreement between them. As
attested by Fig.~\ref{fig: Single impurity unresolved vNEE}(d),
this is true even for a bias voltage that is not relatively small,
such that $\Delta k\approx1$. Recall that the assumption $\Delta k\ll1$
was required only for justifying the approximation of ${\cal C}_{{\rm const}}$,
while both leading terms of the asymptotics are exact regardless of
it.

\subsection{Charge-resolved entanglement\label{subsec:Single-impurity-resolved}}

Next, we studied the symmetry resolution of the R\'enyi moments $Z_{1}$
and $Z_{2}$ and of the vNEE. This was done by extracting the symmetry-resolved
quantities from both the analytical calculation (Eq.~(\ref{eq: Generating function asymptotic expression}))
and the numerical calculation of the generating function $Z_{n}\left(\alpha\right)$,
relying on Eqs.~(\ref{eq: Resolved moment from generating function})
and~(\ref{eq: Resolved vNEE from resolved moment}). The numerical
estimation of $Z_{n}\left(\alpha\right)$ was obtained using Eq.~(\ref{eq: Generating function exact}),
again using the approximation in Eq.~(\ref{eq: Correlation matrix approx})
for the correlation matrix, which assumes an infinite distance between
subsystem $A$ and the impurity.

\begin{figure}[t]
\begin{centering}
\includegraphics[viewport=100bp 25bp 1250bp 565bp,clip,scale=0.36]{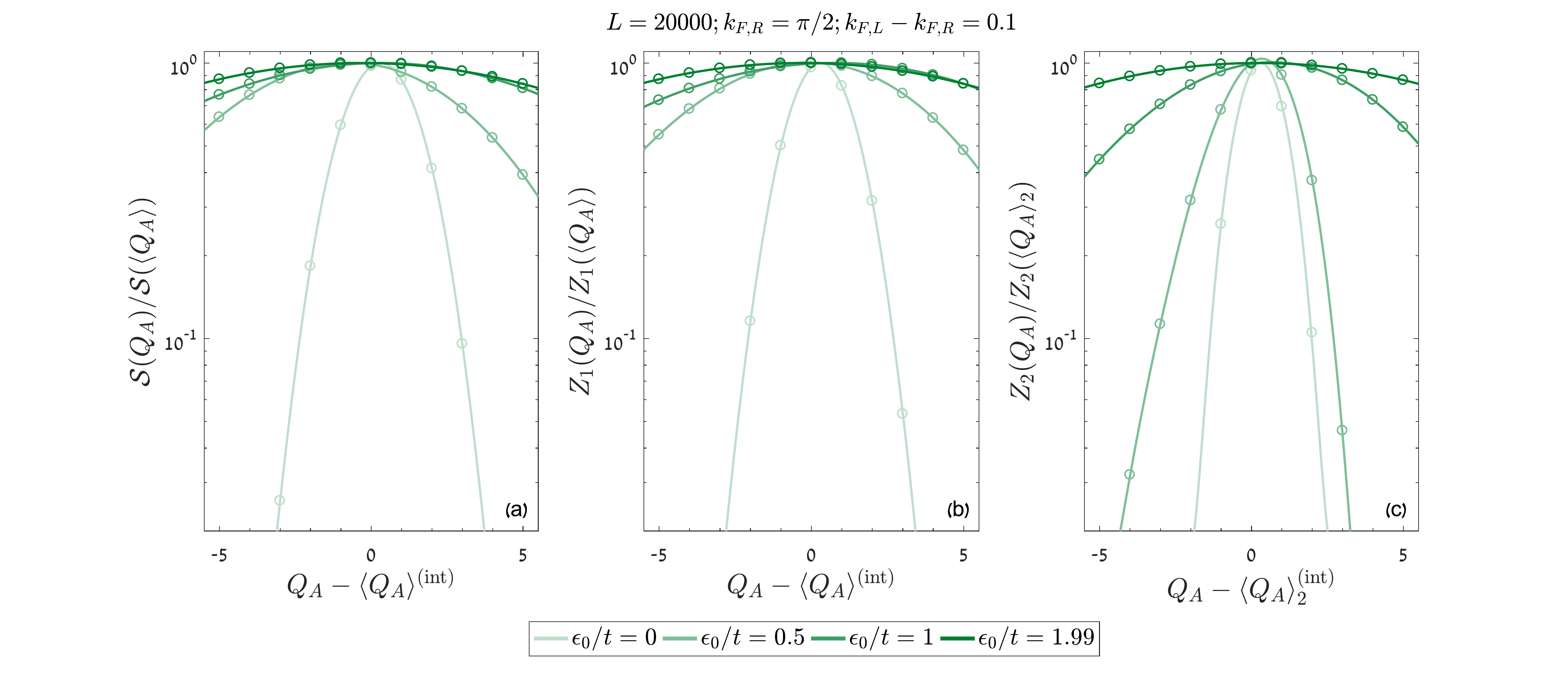}
\par\end{centering}
\caption{\label{fig: Symmetry-resolved-quantities }The single impurity model:
Symmetry-resolved entanglement measures, calculated using Eqs.~(\ref{eq: Resolved moment from generating function})
and~(\ref{eq: Resolved vNEE from resolved moment}), where the generating
function was evaluated analytically through Eq.~(\ref{eq: Generating function asymptotic expression})
(lines) and numerically through Eqs.~(\ref{eq: Correlation matrix approx})
and~(\ref{eq: Generating function exact}) (circles), in a subsystem
of $L=20000$ sites. Results are shown as a function of $Q_{A}$,
the charge in subsystem $A$, for various fixed values of $\epsilon_{0}/t$,
with $k_{F,R}=\pi/2$ and $k_{F,L}-k_{F,R}=0.1$. The charge $Q_{A}$
is measured relative to the rounded mean charge $\left\langle Q_{A}\right\rangle _{n}^{\left({\rm int}\right)}$
defined in Eq.~(\ref{eq: Rounded mean charge}), and the results
are normalized by the analytical value exactly at $\left\langle Q_{A}\right\rangle _{n}$.
Panels (a) and (b) show the resolved vNEE and first R\'enyi moment,
respectively, for which we set $n=1$ (with $\left\langle Q_{A}\right\rangle =\left\langle Q_{A}\right\rangle _{n=1}$),
whereas panel (c) shows the resolved second R\'enyi moment, where
we use $n=2$.}
\end{figure}

The results are presented in Fig.~\ref{fig: Symmetry-resolved-quantities },
where it is evident that the numerical results (naturally sampled
at integer values of $Q_{A}$) accurately fit the analytical results
near the mean charges of the distributions, given by $\left\langle Q_{A}\right\rangle _{n}$
from Eq.~(\ref{eq: Generalized mean charge}) (with $n=1$ for ${\cal S}\left(Q_{A}\right)$
and $Z_{1}\left(Q_{A}\right)$, further simplified in Eq.~(\ref{eq: Mean charge shift}),
and with $n=2$ for $Z_{2}\left(Q_{A}\right)$). The plots in Fig.~\ref{fig: Symmetry-resolved-quantities }
are all centered around the integer charge that is the nearest to
the corresponding mean charge, given by
\begin{equation}
\left\langle Q_{A}\right\rangle _{n}^{\left({\rm int}\right)}=\bigg\lceil\frac{1}{2}\big\lfloor2\left\langle Q_{A}\right\rangle _{n}\big\rfloor\bigg\rceil,\label{eq: Rounded mean charge}
\end{equation}
where $\lfloor\rfloor$ is the floor function ($\lfloor x\rfloor$
is the nearest integer to $x$ from below) and $\lceil\rceil$ is
the ceiling function ($\lceil x\rceil$ is the nearest integer to
$x$ from above). The charge-resolved quantities indeed reach their
maximal value near the mean charge $\left\langle Q_{A}\right\rangle _{n}$,
but due to their slight deviation from a Gaussian form (since $L$
is large but finite) the charge sector $Q_{A}=\left\langle Q_{A}\right\rangle _{n}^{\left({\rm int}\right)}$
is not necessarily the sector where they peak. Note that the mean
charge $\left\langle Q_{A}\right\rangle _{n}$ varies with the model
parameters $\epsilon_{0}/t$ and $k_{\pm}$.

A conspicuous property of the resolved quantities is that their distribution
among charge sectors becomes wider as the model parameters approach
values such that $\left|t_{L}\left(k\right)\right|^{2}\approx\left|r_{R}\left(k\right)\right|^{2}$
for $k_{-}<k<k_{+}$. This is manifested in the analytical results
most simply for $Z_{1}\left(Q_{A}\right)$, since the leading ${\cal O}\left(L\right)$
term in the analytical expression for the charge variance in Eq.~(\ref{eq: Charge variance shift})
peaks exactly within that region in the space of the model parameters.
In Fig.~\ref{fig: Symmetry-resolved-quantities } we fixed $k_{-}=\pi/2$
and $\Delta k=0.1$, so this condition is equivalent there to $\epsilon_{0}\approx2t$.
The exact point $\epsilon_{0}=2t$ cannot be examined analytically
due to a non-integrable divergence of the generating function, as
explained in Subsec.~\ref{subsec:Subleading-asymptotics-of}. Nevertheless,
the results for $\epsilon_{0}=1.99t$ in Fig.~\ref{fig: Symmetry-resolved-quantities }
indicate that, even at points very close to the specific point where
the divergence occurs, this divergence does not cause any discernible
deviation of the analytical calculation from numerical results. 

\begin{figure}[t]
\begin{centering}
\includegraphics[viewport=140bp 20bp 1280bp 570bp,clip,scale=0.38]{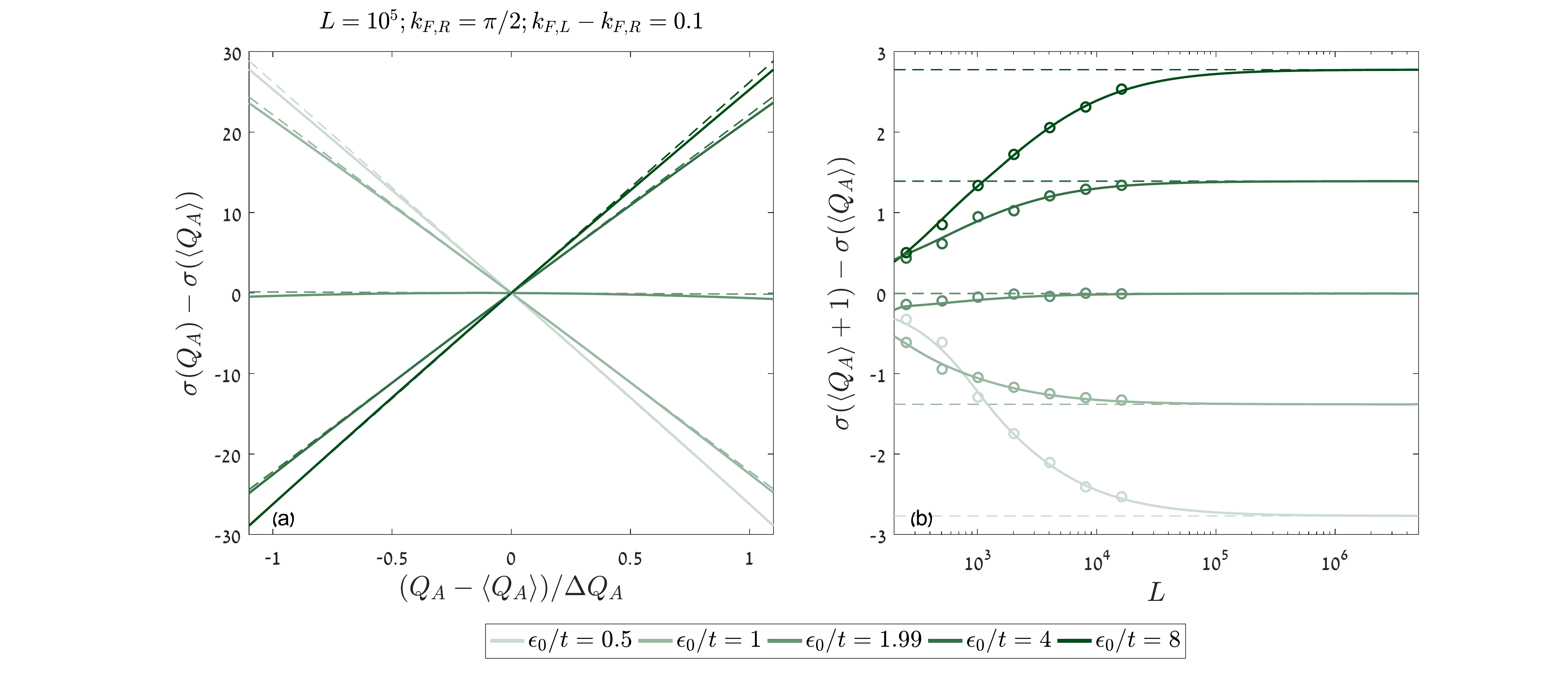}
\par\end{centering}
\caption{\label{fig: Equipartition breaking}The single impurity model: Breaking
of entanglement equipartition by the post-projection vNEE $\sigma\left(Q_{A}\right)$,
calculated using Eqs.~(\ref{eq: Definition for post measurement vNEE}),
(\ref{eq: Resolved moment from generating function}) and~(\ref{eq: Resolved vNEE from resolved moment}).
Results are shown for various fixed values of $\epsilon_{0}/t$, with
$k_{F,R}=\pi/2$ and $k_{F,L}-k_{F,R}=0.1$. In (a), $\sigma\left(Q_{A}\right)$
is calculated based on the analytically estimated generating function
of Eq.~(\ref{eq: Generating function asymptotic expression}), and
measured relative to its estimated value at the mean charge $\left\langle Q_{A}\right\rangle $
for a subsystem of $L=10^{5}$ sites (solid lines). Dashed lines designate
the corresponding linear equipartition-breaking term that was approximated
analytically for large $L$ in Eq.~(\ref{eq: Equipartition breaking slope}).
In (b) we show the difference in $\sigma\left(Q_{A}\right)$ due to
a shift of $Q_{A}$ in one charge unit near the mean charge $\left\langle Q_{A}\right\rangle $
as a function of the length $L$ of subsystem $A$. $\sigma\left(Q_{A}\right)$
is calculated both based on the analytically estimated generating
function of Eq.~(\ref{eq: Generating function asymptotic expression})
(solid lines), and based on the numerically estimated generating function
of Eqs.~(\ref{eq: Correlation matrix approx}) and~(\ref{eq: Generating function exact})
(circles; in the numerical case, $\sigma$ is estimated at the genuine
charge sectors $\left\langle Q_{A}\right\rangle ^{\left({\rm int}\right)}$
and $\left\langle Q_{A}\right\rangle ^{\left({\rm int}\right)}+1$
rather than at $\left\langle Q_{A}\right\rangle $ and $\left\langle Q_{A}\right\rangle +1$).
Dashed lines designate the $L$-independent slope of the linear equipartition-breaking
term in Eq.~(\ref{eq: Equipartition breaking slope}).}
\end{figure}

Additionally, we examined the post-projection charge-resolved vNEE,
$\sigma\left(Q_{A}\right)$, for the single impurity model. Since
the analytical result of Eq.~(\ref{eq: Result for post measurement vNEE})
relies on the Gaussian approximation of the generating function, it
was natural to test whether it properly captured the behavior of the
charge-resolved measures that were extracted from the more accurate
analytical form of the generating function, given by Eq.~(\ref{eq: Generating function asymptotic expression}).
In Fig.~\ref{fig: Equipartition breaking} we present a comparison
between the deviation from entanglement equipartition of the full
analytical result and the leading-order term estimated in Eq.~(\ref{eq: Equipartition breaking slope}),
which is linear in $Q_{A}-\left\langle Q_{A}\right\rangle $. From
Fig.~\ref{fig: Equipartition breaking}(a) it is evident that the
linear breaking of entanglement equipartition holds up to $\left|Q_{A}-\left\langle Q_{A}\right\rangle \right|\approx\Delta Q_{A}$,
confirming that the equipartition-breaking term scales as $\sqrt{L}$
for large $L$. Fig.~\ref{fig: Equipartition breaking}(b) affirms
that for large $L$, the slope of this linear term becomes independent
of $L$.

Fig.~\ref{fig: Equipartition breaking}(b) also includes fully numerical
estimations (using the numerical calculation of $Z_{n}\left(\alpha\right)$
and Eqs.~(\ref{eq: Definition for post measurement vNEE}), (\ref{eq: Resolved moment from generating function})
and~(\ref{eq: Resolved vNEE from resolved moment})) of the change
in $\sigma\left(Q_{A}\right)$ between adjacent (integer valued) charge
sectors near $\left\langle Q_{A}\right\rangle $, for reasonable subsystem
sizes. These numerical results nicely follow the trend of their analytical
counterparts, once again attesting to the validity of the latter. 

\subsection{Accuracy of the generating function calculation\label{subsec:Accuracy-of-analytics}}

\begin{figure}[t]
\begin{centering}
\includegraphics[viewport=170bp 50bp 1190bp 565bp,clip,scale=0.4]{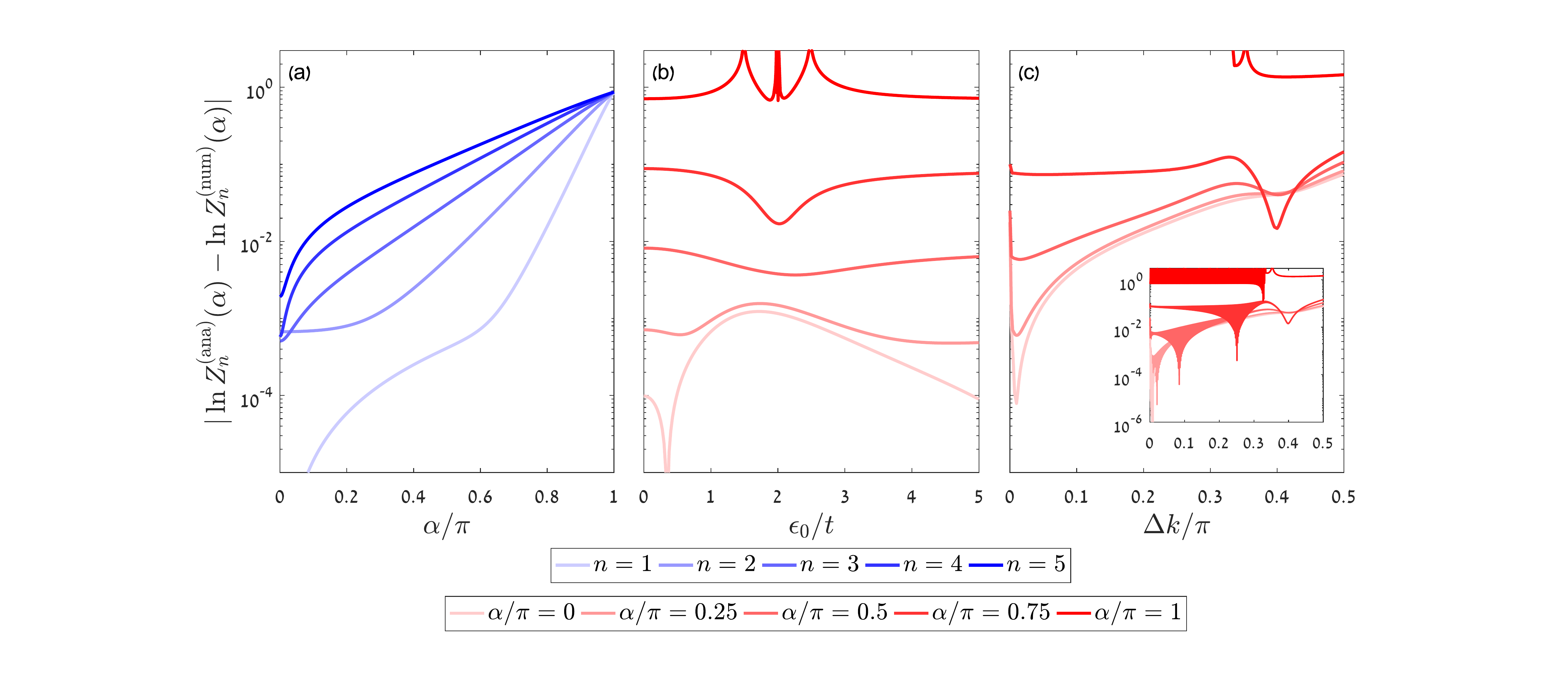}
\par\end{centering}
\caption{\label{fig: Analytical-Numerical Comparison}The single impurity model:
The absolute deviation of the analytical result $\ln Z_{n}^{{\rm (ana)}}\left(\alpha\right)$,
given by Eq.~(\ref{eq: Generating function asymptotic expression}),
from the numerical result $\ln Z_{n}^{{\rm (num)}}\left(\alpha\right)$
calculated using Eqs.~(\ref{eq: Correlation matrix approx}) and~(\ref{eq: Generating function exact}),
in a subsystem of $L=1000$ sites. In the different panels, the absolute
deviation is presented (a) as a function of $\alpha$ for various
fixed values of $n$, with $\epsilon_{0}=t$, $k_{F,R}=\pi/2$ and
$k_{F,L}-k_{F,R}=0.1$; (b) as a function of $\epsilon_{0}/t$ for
various fixed values of $\alpha$, with $n=2$, $k_{F,R}=\pi/2$ and
$k_{F,L}-k_{F,R}=0.1$; and (c) as a function of $\Delta k=k_{F,L}-k_{F,R}$
for various fixed values of $\alpha$, with $n=2$, $\epsilon_{0}=t$
and $k_{F,R}=\pi/2$. In (c) the results oscillate rapidly with varying
$\Delta k$, so only the top envelope of these oscillations (designating
the local maximum of the deviation) is shown, while the inset shows
the full oscillations (for $\alpha=\pi$ there is a regime of $\Delta k$
for which at each period of the oscillation there is a point where
$Z_{n}^{{\rm (num)}}\left(\alpha=\pi\right)$ vanishes while $Z_{n}^{{\rm (ana)}}\left(\alpha=\pi\right)$
does not, and therefore the top envelope is infinite and is not shown).}
\end{figure}

With the generating function $Z_{n}\left(\alpha\right)$ being the
basis for all the analytical calculations discussed in this paper,
a test of the accuracy of its calculation across the parameter space
is required. In Fig.~\ref{fig: Analytical-Numerical Comparison}
the analytical estimation of $\ln Z_{n}\left(\alpha\right)$ (denoted
as $\ln Z_{n}^{\left({\rm ana}\right)}\left(\alpha\right)$) for the
single impurity model is compared to a numerical calculation of $\ln Z_{n}\left(\alpha\right)$
(denoted as $\ln Z_{n}^{\left({\rm num}\right)}\left(\alpha\right)$).
Here the numerical calculation once again neglects the effects of
a finite distance between subsystem $A$ and the impurity (whose effects
will be examined later on), relying on Eqs.~(\ref{eq: Correlation matrix approx})
and~(\ref{eq: Generating function exact}). The comparison indicates
good agreement between analytical and numerical results for values
of $\alpha$ far enough from $\alpha=\pm\pi$. As explained above,
the fact that this agreement breaks down as $\alpha$ approaches $\pm\pi$
is a well-known trait of the leading-order approximation that stems
from the Fisher-Hartwig conjecture~\citep{Bonsignori_2019,Fraenkel_2020,Capizzi_2020}.

Importantly, Fig.~\ref{fig: Analytical-Numerical Comparison} illustrates
that for fixed values of $k_{\pm}$, divergences in the absolute deviation
between analytical and numerical results may occur at $\alpha=\pm\pi$,
though they are typically rare within the parameter space. Such singularities
appear if either $\nu\left(k_{-}\right)=0$, $\nu\left(k_{+}\right)=0$
or $\nu_{0}=0$, where $\ln Z_{n}^{\left({\rm ana}\right)}\left(\alpha=\pm\pi\right)$
diverges (as discussed in Subsec.~\ref{subsec:Subleading-asymptotics-of},
see Eq.~(\ref{eq: Divergence at nu=00003D0})), or if one of the
numerically calculated eigenvalues $\nu_{l}$ vanishes, thus setting
the numerical result to $Z_{n}^{\left({\rm num}\right)}\left(\alpha=\pm\pi\right)=0$
according to Eq.~(\ref{eq: Generating function exact}). In Fig.~\ref{fig: Analytical-Numerical Comparison}(b),
for example, singularities at $\alpha=\pi$ may be detected for values
of $\epsilon_{0}$ near $2t$ (divergence of $\ln Z_{n}^{\left({\rm ana}\right)}\left(\alpha\right)$)
and near $1.5t,2.5t$ (points where $Z_{n}^{\left({\rm num}\right)}\left(\alpha\right)$
vanishes; the locations of these points in the parameter space varies
with $L$). However, when varying $k_{\pm}$ and fixing all other
parameters, $\ln Z_{n}^{\left({\rm num}\right)}\left(\alpha\right)$
oscillates rapidly as a function of $\Delta k$, with periodicity
$2\pi/L$. These oscillations are not reflected in the analytical
result, causing its deviation from $\ln Z_{n}^{\left({\rm num}\right)}\left(\alpha\right)$
to oscillate rapidly as well, as seen in Fig.~\ref{fig: Analytical-Numerical Comparison}(c).
Fig.~\ref{fig: Analytical-Numerical Comparison}(c) more specifically
indicates that for $\alpha=\pm\pi$, there exists a regime of values
of $k_{\pm}$ where in each period of the oscillation there is a singularity
of the deviation. Each such singularity corresponds to a vanishing
numerically calculated eigenvalue $\nu_{l}$, i.e.~to a zero of $Z_{n}^{\left({\rm num}\right)}\left(\alpha=\pm\pi\right)$.

\begin{figure}[t]
\begin{centering}
\includegraphics[viewport=80bp 0bp 1240bp 585bp,clip,scale=0.36]{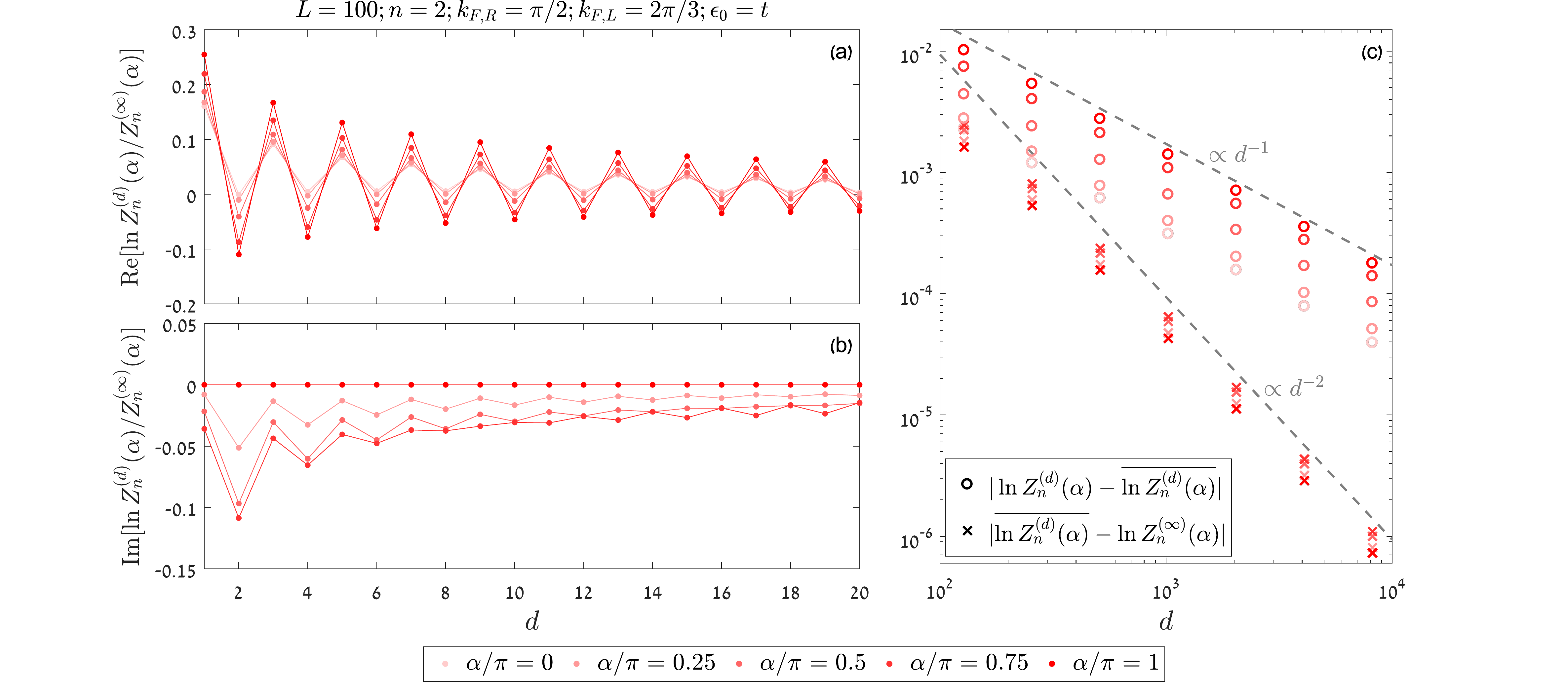}
\par\end{centering}
\caption{\label{fig: Hankel term}The single impurity model: The deviation
of $\ln Z_{n}^{\left(d\right)}\left(\alpha\right)$, the distance-dependent
generating function calculated numerically using Eqs.~(\ref{eq: Correlation matrix full expression})
and~(\ref{eq: Generating function exact}) (where $d$ is the distance
between the leftmost site of $A$ and the origin, at which the scatterer
is located), from $\ln Z_{n}^{\left(\infty\right)}\left(\alpha\right)$,
the generating function for $d\to\infty$ calculated numerically using
Eqs.~(\ref{eq: Correlation matrix approx}) and~(\ref{eq: Generating function exact}),
in a subsystem of $L=100$ sites. Results are shown for various fixed
values of $\alpha$, with $n=2$, $\epsilon_{0}=t$, $k_{F,R}=\pi/2$
and $k_{F,L}=2\pi/3$. Both (a) the real part and (b) the imaginary
part of the deviation are plotted (dots), accompanied by thin lines
as guides to the eye (note that for $\alpha=0$ the imaginary part
vanishes by definition). Panel (c) shows the deviation for $d>L$
following averaging over oscillations, where the absolute values of
both the average $\overline{\ln Z_{n}^{\left(d\right)}\left(\alpha\right)}-\ln Z_{n}^{\left(\infty\right)}\left(\alpha\right)$
and the amplitude $\ln Z_{n}^{\left(d\right)}\left(\alpha\right)-\overline{\ln Z_{n}^{\left(d\right)}\left(\alpha\right)}$
of the oscillations are plotted. Dashed gray lines emphasize that
for all plotted values of $\alpha$, the average deviation approaches
a $\propto d^{-2}$ power law behavior and the amplitude approaches
a $\propto d^{-1}$ power law behavior.}
\end{figure}

Finally, we address the effect of a finite distance between subsystem
$A$ and the impurity -- i.e., of the Hankel term in Eq.~(\ref{eq: Correlation matrix full expression})
that was omitted from Eq.~(\ref{eq: Correlation matrix approx}),
and was heretofore disregarded. Although we were not able to incorporate
the effect of the Hankel term into our analytical calculation, the
concrete example of the single impurity model allows us to examine
numerically the dependence of the results on $d$, the distance of
$A$ from the impurity at the origin. More precisely, $d$ is taken
to be the location of the leftmost site in subsystem $A$, such that
$A$ includes the sites $n=d,\ldots,d+L-1$. Fig.~\ref{fig: Hankel term}
shows the comparison between two numerical calculations of the generating
function: the one extracted from the approximate form of the correlation
matrix in Eq.~(\ref{eq: Correlation matrix approx}) (denoted as
$Z_{n}^{\left(\infty\right)}\left(\alpha\right)$), which was used
in the comparison to the analytical results, and the one that relies
on the full form of the correlation matrix in Eq.~(\ref{eq: Correlation matrix full expression})
(denoted as $Z_{n}^{\left(d\right)}\left(\alpha\right)$).

One may observe that the additional term that accounts for finite
$d$ effects on $\ln Z_{n}\left(\alpha\right)$ oscillates as a function
of $d$, a manifestation of Friedel oscillations~\citep{Friedel1958,coleman_2015}.
The typical wavenumber of these oscillations is $2k_{F,R}$, and is
therefore independent of $k_{F,L}$, as should be expected from the
fact that the $d$-dependent term in Eq.~(\ref{eq: Correlation matrix full expression})
is independent of $k_{F,L}$ as well (this holds for a subsystem to
the right of the scattering region; for a subsystem on the left, the
roles of $k_{F,L}$ and $k_{F,R}$ are switched). The numerical results
also verify that the deviation of $\ln Z_{n}^{\left(d\right)}\left(\alpha\right)$
from $\ln Z_{n}^{\left(\infty\right)}\left(\alpha\right)$ indeed
vanishes as $d\to\infty$. Furthermore, when averaging over the oscillations,
one finds that for $d\gg L$ their average approaches a power law
decay proportional to $d^{-2}$, while their amplitude exhibits a
power law behavior proportional to $d^{-1}$, typical of Friedel oscillations
in 1D\footnote{In $D$ dimensions, Friedel oscillations tend to decay as $1/R^{D}$,
where $R$ is the distance from the impurity.}~\citep{Friedel1958}. This observation dovetails with the aforementioned
projection of an algebraic decay of the Hankel term in the correlation
matrix.

The dependence of the generating function on $d$ may be seen as an
effect of boundary conditions, referring here to the boundary of subsystem
$A$. Finite $d$ effects are therefore expected not to be reflected
in the leading, linear in $L$ term of $\ln Z_{n}\left(\alpha\right)$,
since this term represents an extensive property of $A$. We have
verified numerically that the Hankel contribution indeed has no extensive
effect. The logarithmic term of $\ln Z_{n}\left(\alpha\right)$, in
contrast, does generally depend on the boundary conditions of $A$
(cf.~Refs.~\citep{PhysRevLett.96.100503,PhysRevLett.112.160403}),
and so for finite $d$ the Hankel contribution may be proportional
to $\ln L$, though such a logarithmic dependence is in practice hard
to ascertain numerically for accessible subsystem sizes. Regardless
of this, considering that in the previous calculations we kept terms
which are constant in $L$, and that the Hankel contribution is proportional
to $d^{-1}$ provided that $d\gg L$, neglecting the $d$ dependence
is certainly justified in the regime $d\gg L$.

\section{Generalization to multiple scatterers\label{sec:Generalization-to-multiple}}

In the following section we rely on the analytical results for the
model described in Sec.~\ref{sec: The Physical Model} in order to
derive corresponding results for a more general scenario, where the
tight-binding chain contains several different scattering regions
rather than just a single one. The necessary foundation is the description
of the combined scattering effects of two scatterers, with a distance
of $\ell$ sites between them. We assume throughout this section that
$\ell$ is considerably larger than the Fermi wavelengths, $\left(k_{F,L}\right)^{-1}$
and $\left(k_{F,R}\right)^{-1}$. We mark the left scatterer with
${\rm I}$, and the right one with ${\rm II}$.

\begin{figure}[t]
\begin{centering}
\includegraphics[viewport=180bp 20bp 1100bp 710bp,clip,scale=0.35]{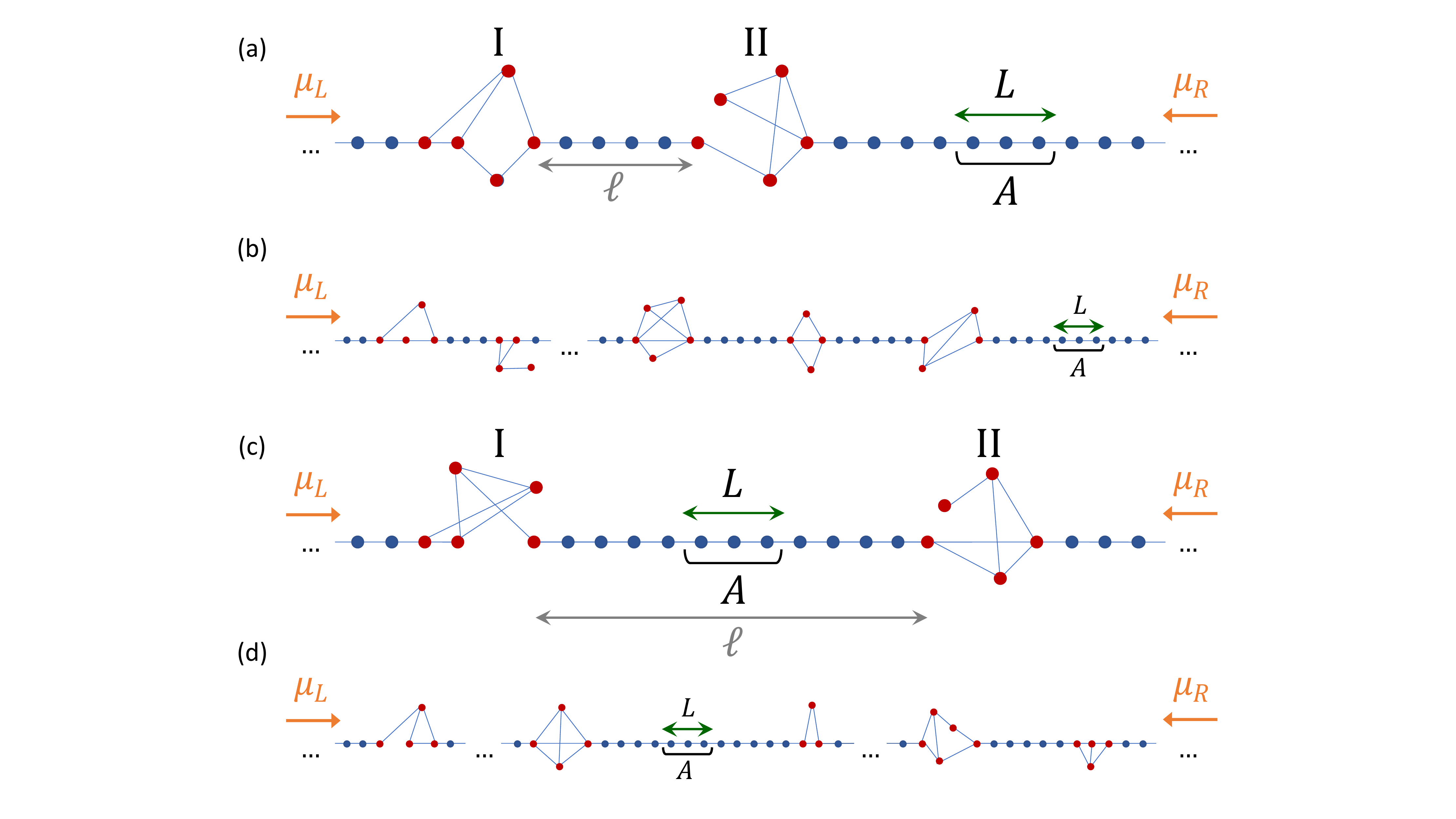}
\par\end{centering}
\caption{\label{fig:Schematic-multiple}Schematic illustrations of the lattice
models containing multiple scattering regions, which are discussed
throughout Sec.~\ref{sec:Generalization-to-multiple}. Sites marked
in blue belong to the unperturbed parts of the tight-binding chain,
while sites marked in red belong to the scattering regions. $\mu_{L}$
($\mu_{R}$) designates the chemical potential for particles incoming
from the left (right). $A$ denotes the subsystem of $L$ contiguous
sites with respect to which the calculations of bipartite entanglement
are performed. Panel (a) depicts a scenario where two scattering regions
are located on the same side of $A$. Panel (b) depicts a more general
scenario with multiple scattering regions, that are all on the same
side of $A$. Panel (c) depicts a scenario where $A$ is located between
two scattering regions. Panel (d) depicts an example for the general
scenario, with multiple scattering regions on both sides of $A$.
In (a) and (c), $\ell$ denotes the number of sites between scattering
regions ${\rm I}$ and ${\rm II}$, which is assumed to be much larger
than the Fermi wavelengths.}
\end{figure}

Let us associate a unitary scattering matrix with each scattering
region,
\begin{equation}
S_{i}\left(k\right)=\left(\begin{array}{cc}
r_{L}^{\left(i\right)}\left(k\right) & t_{R}^{\left(i\right)}\left(k\right)\\
t_{L}^{\left(i\right)}\left(k\right) & r_{R}^{\left(i\right)}\left(k\right)
\end{array}\right),
\end{equation}
where $i={\rm I},{\rm II}$. If subsystem $A$ is situated to the
same side of both scattering regions, as depicted in Fig.~\ref{fig:Schematic-multiple}(a),
we may treat them as a single scatterer with appropriate reflection
and transmission amplitudes~\citep{datta_1995}. For a wave incoming
from the left, these are given by
\begin{align}
r_{L}^{\left({\rm I}+{\rm II}\right)} & =r_{L}^{\left({\rm I}\right)}+\frac{t_{R}^{\left({\rm I}\right)}t_{L}^{\left({\rm I}\right)}r_{L}^{\left({\rm II}\right)}}{e^{-2ik\ell}-r_{R}^{\left({\rm I}\right)}r_{L}^{\left({\rm II}\right)}},\,\,\,\,\,t_{L}^{\left({\rm I}+{\rm II}\right)}=\frac{t_{L}^{\left({\rm I}\right)}t_{L}^{\left({\rm II}\right)}}{1-e^{2ik\ell}r_{R}^{\left({\rm I}\right)}r_{L}^{\left({\rm II}\right)}},
\end{align}
and for a wave incoming from the right the amplitudes $r_{R}^{\left({\rm I}+{\rm II}\right)}$
and $t_{R}^{\left({\rm I}+{\rm II}\right)}$ are given by the same
expressions, up to exchanging $R\leftrightarrow L,{\rm I}\leftrightarrow{\rm II}$
and multiplying by an overall phase (the notation emphasizing the
dependence of the scattering amplitudes on $k$ has been omitted for
brevity).

As described in Sec.~\ref{sec:Analytical-Asymptotics}, the correlation
matrix $C_{A}$ is the basis for our calculation of entanglement measures.
By the same line of argument of Sec.~\ref{sec: The Physical Model},
it can be written as a sum of a Toeplitz matrix and a Hankel matrix,
where the Hankel matrix can be neglected assuming $A$ is far enough
from the combined scatterer, as before. For the estimation of the
Toeplitz term, we define the following incoherent scattering probabilities~\citep{datta_1995}:
\begin{align}
\left|\overline{r}_{L}\right|^{2} & =\left|r_{L}^{\left({\rm I}\right)}\right|^{2}+\frac{\left|t_{R}^{\left({\rm I}\right)}t_{L}^{\left({\rm I}\right)}r_{L}^{\left({\rm II}\right)}\right|^{2}}{1-\left|r_{R}^{\left({\rm I}\right)}r_{L}^{\left({\rm II}\right)}\right|^{2}},\,\,\,\,\,\left|\overline{t}_{L}\right|^{2}=\frac{\left|t_{L}^{\left({\rm I}\right)}t_{L}^{\left({\rm II}\right)}\right|^{2}}{1-\left|r_{R}^{\left({\rm I}\right)}r_{L}^{\left({\rm II}\right)}\right|^{2}}.\label{eq:Incoherent-probabilities}
\end{align}
The probabilities $\left|\overline{r}_{L}\left(k\right)\right|^{2}$
and $\left|\overline{t}_{L}\left(k\right)\right|^{2}$ differ from
the expressions $\left|r_{L}^{\left({\rm I}+{\rm II}\right)}\left(k\right)\right|^{2}$
and $\left|t_{L}^{\left({\rm I}+{\rm II}\right)}\left(k\right)\right|^{2}$
(respectively) by terms that oscillate as $e^{2ik\ell}$, and that
(recalling the assumption $\ell\gg\left(k_{F,L}\right)^{-1},\left(k_{F,R}\right)^{-1}$)
can thus be neglected when integrating over $k$. Analogous definitions
of probabilities $\left|\overline{r}_{R}\left(k\right)\right|^{2}$
and $\left|\overline{t}_{R}\left(k\right)\right|^{2}$ (again applying
the replacements $R\leftrightarrow L,{\rm I}\leftrightarrow{\rm II}$
in Eq.~(\ref{eq:Incoherent-probabilities})) are used to replace
$\left|r_{R}^{\left({\rm I}+{\rm II}\right)}\left(k\right)\right|^{2}$
and $\left|t_{R}^{\left({\rm I}+{\rm II}\right)}\left(k\right)\right|^{2}$,
respectively, under integrals over $k$. This finally allows us to
write the two-point correlation matrix of $A$ according to Eq.~(\ref{eq: Correlation matrix approx}),
where the incoherent scattering probabilities stand in for the original
scattering probabilities of a single scattering region. This scheme
can be readily extended to treat multiple scattering regions that
are all located on the same side of subsystem $A$ (as illustrated
in Fig.~\ref{fig:Schematic-multiple}(b)) as a single combined scattering
region, and therefore one may use the analytical results of Sec.~\ref{sec:Analytical-Asymptotics}
to estimate the various entanglement measures discussed there.

Considering the case where scattering regions are located on both
sides of subsystem $A$ (depicted generally in Fig.~\ref{fig:Schematic-multiple}(d)),
this scheme of calculating combined scattering probabilities has reduced
the problem to that of two scattering regions -- region ${\rm I}$
to the left of $A$, and region ${\rm II}$ to its right, as illustrated
in Fig.~\ref{fig:Schematic-multiple}(c). The distances from the
edges of $A$ to the scatterers are assumed to be much larger than
$L$, the number of sites in $A$. Using the assumption of the distance
between scatterers ${\rm I}$ and ${\rm II}$ being much larger than
the Fermi wavelengths, we arrive at the following approximation of
the correlation matrix in that region:
\begin{equation}
C_{mn}\approx\frac{1}{2\pi}\underset{-\pi}{\overset{\pi}{\int}}e^{-i\left(m-n\right)k}\overline{\tau}\left(k\right)dk.\label{eq: Correlation matrix between two scatterers}
\end{equation}
Here we have defined
\begin{equation}
\overline{\tau}\left(k\right)=\begin{cases}
1 & -k_{-}<k<k_{-},\\
\frac{1}{2}\left(1+\overline{\nu}_{{\rm I}}\left(k\right)\right) & k_{-}<k<k_{+},\\
\frac{1}{2}\left(1-\overline{\nu}_{{\rm II}}\left(-k\right)\right) & -k_{+}<k<-k_{-},\\
0 & \text{\text{otherwise}},
\end{cases}\label{eq: Toeplitz symbol between two scatterers}
\end{equation}
where
\begin{equation}
\frac{1\pm\overline{\nu}_{{\rm I}}\left(k\right)}{2}=\frac{\left|t_{L}^{\left({\rm I}\right)}\left(k\right)\right|^{2}}{1-\left|r_{R}^{\left({\rm I}\right)}\left(k\right)r_{L}^{\left({\rm II}\right)}\left(k\right)\right|^{2}}\,\,\,\,\,\text{for }k_{F,L}=k_{\pm},
\end{equation}
and $\overline{\nu}_{{\rm II}}\left(k\right)$ is defined similarly,
up to replacing $\left|t_{L}^{\left({\rm I}\right)}\left(k\right)\right|^{2}$
with $\left|t_{R}^{\left({\rm II}\right)}\left(k\right)\right|^{2}$.
The derivation of Eq.~(\ref{eq: Correlation matrix between two scatterers})
is detailed in Appendix~\ref{subsec:Correlation-matrix-two-scatterers}.

The analytical method of Subsec.~\ref{subsec: Leading analytical asymptotics}
can now be applied to obtain the exact form of the two leading terms
in the asymptotic expression for the generating function defined in
Eq.~(\ref{eq: Generating function definition}):
\begin{align}
\ln Z_{n}\left(\alpha\right) & \sim\frac{L}{2\pi}\left[2i\alpha k_{-}+\underset{k_{-}}{\overset{k_{+}}{\int}}\left\{ e_{n}^{\left(\alpha\right)}\left(1,\overline{\nu}_{{\rm I}}\left(k\right)\right)+e_{n}^{\left(\alpha\right)}\left(1,-\overline{\nu}_{{\rm II}}\left(k\right)\right)\right\} dk\right]\nonumber \\
 & +\ln L\left[Q_{n}\left(\overline{\nu}_{{\rm I}}\left(k_{-}\right),\alpha\right)+Q_{n}\left(-\overline{\nu}_{{\rm I}}\left(k_{+}\right),-\alpha\right)\right]\nonumber \\
 & +\ln L\left[Q_{n}\left(-\overline{\nu}_{{\rm II}}\left(k_{-}\right),\alpha\right)+Q_{n}\left(\overline{\nu}_{{\rm II}}\left(k_{+}\right),-\alpha\right)\right].\label{eq: GenFun Asymptotics between two scatterers}
\end{align}
An approximation for the first subleading correction to this asymptotics,
which is independent of $L$, may be derived by following the analytical
method of Subsec.~\ref{subsec:Subleading-asymptotics-of}. The various
entanglement measures that were discussed in the context of a single
scattering region can now be extracted from Eq.~(\ref{eq: GenFun Asymptotics between two scatterers}).

We report in particular the result for the unresolved vNEE in subsystem
$A$ between the two scatterers. For convenience, we introduce the
notations
\begin{equation}
{\cal T}_{{\rm I}}\left(k\right)=\frac{\left|t_{L}^{\left({\rm I}\right)}\left(k\right)\right|^{2}}{1-\left|r_{R}^{\left({\rm I}\right)}\left(k\right)r_{L}^{\left({\rm II}\right)}\left(k\right)\right|^{2}}\,\,\,\,\,,\,\,\,\,\,{\cal T}_{{\rm II}}\left(k\right)=\frac{\left|t_{R}^{\left({\rm II}\right)}\left(k\right)\right|^{2}}{1-\left|r_{R}^{\left({\rm I}\right)}\left(k\right)r_{L}^{\left({\rm II}\right)}\left(k\right)\right|^{2}}.
\end{equation}
The vNEE is given by the asymptotic form
\begin{align}
{\cal S} & \sim-\frac{L}{2\pi}\underset{k_{-}}{\overset{k_{+}}{\int}}\left[{\cal T}_{{\rm I}}\left(k\right)\ln{\cal T}_{{\rm I}}\left(k\right)+\left(1-{\cal T}_{{\rm I}}\left(k\right)\right)\ln\left(1-{\cal T}_{{\rm I}}\left(k\right)\right)\right]dk\nonumber \\
 & -\frac{L}{2\pi}\underset{k_{-}}{\overset{k_{+}}{\int}}\left[{\cal T}_{{\rm II}}\left(k\right)\ln{\cal T}_{{\rm II}}\left(k\right)+\left(1-{\cal T}_{{\rm II}}\left(k\right)\right)\ln\left(1-{\cal T}_{{\rm II}}\left(k\right)\right)\right]dk\nonumber \\
 & +\left[q\left({\cal T}_{{\rm I}}\left(k_{F,R}\right)\right)+q\left(1-{\cal T}_{{\rm I}}\left(k_{F,L}\right)\right)+q\left({\cal T}_{{\rm II}}\left(k_{F,L}\right)\right)+q\left(1-{\cal T}_{{\rm II}}\left(k_{F,R}\right)\right)\right]\ln L+{\cal O}\left(1\right),\label{eq: Unresolved vNEE multiple scatterers}
\end{align}
where the function $q\left(p\right)$ was defined in Eq.~(\ref{eq: Auxiliary functions vNEE}).
Note that if we require transmission and reflection factors to be
constant functions of $k$, we recreate Eq.~(26) of Ref.~\citep{PhysRevB.96.054302}.
We have therefore generalized the scenario discussed in Ref.~\citep{PhysRevB.96.054302},
where the subsystem lies on a tight binding chain coupled to two macroscopic
leads with $k$-independent hybridization factors.

\section{Conclusions and outlook\label{sec:Conclusions-and-outlook}}

While the exact physical description of nonequilibrium many-body states
remains a coveted yet elusive goal, entanglement measures continue
to facilitate incremental progress toward its achievement. In this
work we sought to exactly quantify the steady-state entanglement of
a paradigmatic 1D lattice model, namely a homogeneous tight-binding
chain interrupted by an arbitrary number of scattering regions, held
under a bias voltage at zero temperature. For this purpose we employed
the generalized Fisher-Hartwig conjecture to calculate bipartite entanglement
measures for a subsystem located far away from the scatterers.

A central result of our work is given in Eq.~(\ref{eq: Unresolved vNEE asymptotics}).
The von-Neumann entanglement entropy was shown to scale extensively
with the size of the subsystem in question, with an additive logarithmic
correction that arises from the sharp jumps in the energy distribution.
While in the ground state such a scaling law is considered exotic
and requires long range couplings~\citep{Alba_2009,Ares_2014}, the
class of steady states we investigated, which are excited eigenstates
of the Hamiltonian with a Fermi-discontinuous distribution of the
excitations~\citep{Alba_2009}, exhibits it generically. This suggests
that a scaling law of the form of Eq.~(\ref{eq: Unresolved vNEE asymptotics})
should be much more common in nonequilibrium setups, rendering it
a strong signature of the unique properties that distinguish steady
states in and out of equilibrium. More precisely, this entanglement
scaling law should in general be observed in steady states where the
single-particle energy distribution features partially occupied states
(of non-vanishing measure) and Fermi discontinuities. Refs.~\citep{PhysRevB.96.054302,PhysRevA.89.032321}
have indeed found such scaling of the vNEE for nonequilibrium steady
states in some particular impurity setups. We expect this behavior
to apply also to other current-carrying impurity models at zero temperature,
including systems that contain massless Dirac fermions\footnote{Indeed, in the regime of a small bias voltage in our model, the dispersion
relation for states within the voltage window may be linearized, giving
an effective description of them as Dirac fermion states. These states
are those responsible for the emergence of the linear and logarithmic
leading terms in Eq.~(\ref{eq: Unresolved vNEE asymptotics}).} or host dissipative defects~\citep{alba2021noninteracting,chaudhari2021zeno}.

Notably, the form of the extensive term of the vNEE encapsulates the
basic elements defining the steady state. It arises from scattering
states within the energy window between the two different chemical
potentials, as scattered particles traverse the subsystem from side
to side, and thereby entangle its entire bulk to the rest of the chain.
The assumption of a large subsystem allows us to disregard boundary
effects and attribute classical probabilities to the scattering processes,
and therefore the contribution of each mode to the entanglement is
equivalent to a classical mixture entropy. As a consequence, the extensive
term of the vNEE vanishes either in the absence of a bias voltage
(i.e., in equilibrium), or if the scattering region is trivial. In
this sense, the model studied here can be seen as a minimal model
for producing such scaling of the steady state entanglement. The picture
of entanglement as a result of partial occupation of momentum states
due to scattering is also reflected in the linear term of $\ln Z_{n}\left(\alpha\right)$
in Eq.~(\ref{eq: Conjectured entropy asymptotics}), which generalizes
the known full counting statistics formula found by Levitov and Lesovik~\citep{doi:10.1063/1.3149497,levitov1993charge,PhysRevB.75.205329}.

The exact expression for the vNEE is only one out of the comprehensive
set of results presented in this paper, all encoded in the asymptotics
of the generating function $Z_{n}\left(\alpha\right)$ in Eq.~(\ref{eq: Generating function asymptotic expression}).
These results include R\'enyi moments (from which R\'enyi entropies
are readily obtained), statistical charge properties, and charge-resolved
moments and entanglement measures. We particularly emphasize the novelty
of the result in Eq.~(\ref{eq: Result for post measurement vNEE})
for the vNEE following a projective charge measurement $\sigma\left(Q_{A}\right)$.
It implies that to leading order, entanglement is equally distributed
across charge sectors, as has been established for the great majority
of models for which symmetry-resolved entanglement was studied. However,
the term breaking entanglement equipartition grows with the subsystem
size $L$, as $\sigma\left(Q_{A}\right)-{\cal S}={\cal O}\left(\sqrt{L}\right)$,
for charge sectors within the standard deviation from the mean charge.
The leading equipartition-breaking term was also found to be anti-symmetric
in $Q_{A}-\left\langle Q_{A}\right\rangle $.

To the best of our knowledge, the model studied in this paper is the
first where $\sigma\left(Q_{A}\right)$ exhibits these properties.
A natural question for future research is therefore whether this unique
behavior can be exclusively ascribed to the partially-transmitted
current carried by the nonequilibrium steady state, as it was not
witnessed in a different nonequilibrium model where the net current
had been absent~\citep{PhysRevB.103.L041104}. We additionally note
that we expect the new behaviors uncovered in this work to hold in
the presence of interactions, although further investigation is required
to establish this claim.

Our analytical results were tested against numerics in a model where
a single impurity site serves as the scatterer, and were shown to
compare to them favorably. We additionally used numerics to observe
effects of a finite distance between the subsystem and the scattering
region, detecting signatures of Friedel oscillations and confirming
that the effects are indeed negligible when that distance is large
enough. We note that by using various proposed protocols for the measurement
of (resolved and unresolved) entanglement measures~\citep{PhysRevA.97.023604,PhysRevLett.120.050406,vitale2021symmetryresolved,PhysRevA.98.032302,PhysRevA.99.062309},
our results may be experimentally tested in setups based on cold atom
or electronic systems~\citep{datta_1995,Husmann1498}.

The road toward a deeper understanding of nonequilibrium many-body
physics is still riddled with unanswered questions. Hopefully the
exact results presented in this paper can serve as building blocks
for the future description of richer and more intricate phenomena
out of equilibrium, such as effects of interactions, disorder, localization
or external driving~\citep{doi:10.1146/annurev-conmatphys-031214-014726,doi:10.1146/annurev-conmatphys-031214-014701,Lorenzo2017},
entanglement phase transitions~\citep{PhysRevX.5.031032,PhysRevX.5.031033,PhysRevB.100.134306,PhysRevX.9.031009},
and transport through mesoscopic systems~\citep{PhysRevX.9.021007,PhysRevLett.123.110601}.

\section*{Acknowledgments}

We thank P.~Calabrese, M.~Dalmonte and E.~Sela for stimulating
discussions. Our work has been supported by the U.S.-Israel Binational
Science Foundation (Grant No. 2016224).

\appendix

\section*{Appendix}

\section{Detailed derivations \label{sec:Detailed-derivations}}

In what follows we expand on the derivations of central analytical
results discussed in Secs.~\ref{sec:Analytical-Asymptotics} and~\ref{sec:Generalization-to-multiple}.
These analytical results include the expressions for the logarithmic
and constant terms in the asymptotics of $\ln Z_{n}\left(\alpha\right)$
(Appendices~\ref{subsec: Appendix Logarithmic Term} and~\ref{subsec: FH constant term});
the expansion of $\ln Z_{1}\left(\alpha\right)$ in powers of $\alpha$
(Appendix~\ref{subsec: Expansion of generating function}), from
which statistical charge properties were extracted in Subsec.~\ref{subsec:Charge-statistics};
the unresolved vNEE in the presence of a single scatterer (Appendix~\ref{subsec: The von-Neumman-entanglement});
and the generalized form of the correlation matrix for the case of
a subsystem between two scatterers (Appendix~\ref{subsec:Correlation-matrix-two-scatterers}).

\setcounter{equation}{0}
\renewcommand{\theequation}{A.\arabic{equation}}

\subsection{Logarithmic term in $\ln Z_{n}\left(\alpha\right)$\label{subsec: Appendix Logarithmic Term}}

We present here a detailed derivation of the term ${\cal I}_{{\rm log}}\left(n,\alpha\right)$
in the asymptotic form~(\ref{eq: Conjectured entropy asymptotics})
of $\ln Z_{n}\left(\alpha\right)$. Let us start by defining the complex
function $\beta_{a,b}\left(\lambda\right)=\frac{1}{2\pi i}\ln\frac{\lambda-b}{\lambda-a}$
for two real numbers $a<b$, choosing the principal branch of the
logarithm, such that $\left|\mathrm{Re}\beta_{a,b}\left(\lambda\right)\right|<\frac{1}{2}$.
A crucial property of this function is that for a real $x\neq a,b$,
\begin{equation}
\beta_{a,b}\left(x+i0^{\pm}\right)=\begin{cases}
\frac{1}{2\pi i}\ln\left|\frac{x-b}{x-a}\right|\pm\frac{1}{2} & x\in\left(a,b\right),\\
\frac{1}{2\pi i}\ln\left|\frac{x-b}{x-a}\right| & x\notin\left[a,b\right].
\end{cases}\label{eq: Beta on real line}
\end{equation}
Using Eq.~(\ref{eq: Contour integral}) and the asymptotic expression
for $\ln D_{L}\left(\lambda\right)$ in Eq.~(\ref{eq: Conjectured det asymptotics}),
we may write 
\begin{align}
{\cal I}_{{\rm log}}\left(n,\alpha\right) & =\lim_{\varepsilon,\delta\rightarrow0^{+}}\frac{1}{2\pi i}\underset{c\left(\varepsilon,\delta\right)}{\int}\ln L\left(\beta_{\nu\left(k_{-}\right),1}\left(\lambda\right)^{2}+\beta_{-1,\nu\left(k_{+}\right)}\left(\lambda\right)^{2}+\beta_{-1,1}\left(\lambda\right)^{2}\right)\nonumber \\
 & \times\frac{d}{d\lambda}e_{n}^{\left(\alpha\right)}\left(1+\varepsilon,\lambda\right)d\lambda,
\end{align}
where we have employed integration by parts. Using the property from
Eq.~(\ref{eq: Beta on real line}) and taking the limit $\varepsilon,\delta\rightarrow0^{+}$,
we obtain
\begin{equation}
{\cal I}_{{\rm log}}\left(n,\alpha\right)=\frac{1}{2\pi^{2}}\left[\underset{\nu\left(k_{-}\right)}{\overset{1}{\int}}\ln\left|\frac{x-1}{x-\nu\left(k_{-}\right)}\right|+\underset{-1}{\overset{\nu\left(k_{+}\right)}{\int}}\ln\left|\frac{x-\nu\left(k_{+}\right)}{x+1}\right|+\underset{-1}{\overset{1}{\int}}\ln\left|\frac{x-1}{x+1}\right|\right]\frac{d}{dx}e_{n}^{\left(\alpha\right)}\left(1,x\right)dx.
\end{equation}
Through a change of variables $x\rightarrow-x$ we may notice that
\begin{equation}
\underset{-1}{\overset{\nu}{\int}}\ln\left|\frac{x-\nu}{x+1}\right|\frac{d}{dx}e_{n}^{\left(\alpha\right)}\left(1,x\right)dx=\underset{-\nu}{\overset{1}{\int}}\ln\left|\frac{x-1}{x+\nu}\right|\frac{d}{dx}e_{n}^{\left(-\alpha\right)}\left(1,x\right)dx,
\end{equation}
and therefore
\begin{equation}
{\cal I}_{{\rm log}}\left(n,\alpha\right)=Q_{n}\left(\nu\left(k_{-}\right),\alpha\right)+Q_{n}\left(-\nu\left(k_{+}\right),-\alpha\right)+\frac{1}{12}\left(\frac{1}{n}-n\right)-\frac{\alpha^{2}}{4\pi^{2}n}.
\end{equation}
Here we invoked the notation from Eq.~(\ref{eq: Discontinuity kernel notation}),
and the term written explicitly was obtained by carrying out the integration
through the change of variables $u=\ln\left|\frac{x-1}{x+1}\right|$.

\subsection{Subleading term in $\ln Z_{n}\left(\alpha\right)$\label{subsec: FH constant term}}

We detail the calculation of the expression in Eq.~(\ref{eq: FH constant term})
for the term $\tilde{{\cal I}}_{{\rm const}}\left(n,\alpha\right)$,
which is the approximate subleading term in the asymptotics of $\ln Z_{n}\left(\alpha\right)$
in Eq.~(\ref{eq: Generating function asymptotic expression}). The
approximate piecewise-constant symbol $\tilde{\phi}\left(k\right)$
of Eq.~(\ref{eq: Approx Toeplitz symbol}) can be written in a Fisher-Hartwig
form~\citep{10.2307/23030524}:
\begin{equation}
\tilde{\phi}\left(k\right)=E\left(\lambda\right)\cdot e^{i\sum_{j=1}^{3}\left(k-k_{j}\right)\beta_{j}}\prod_{j=1}^{3}g_{j}\left(k\right),
\end{equation}
where $k_{1}\equiv k_{-},\beta_{1}\equiv\beta_{\nu_{0},1}\left(\lambda\right),k_{2}\equiv k_{+},\beta_{2}\equiv\beta_{-1,\nu_{0}}\left(\lambda\right),k_{3}\equiv2\pi-k_{F,R},\beta_{3}\equiv-\beta_{-1,1}\left(\lambda\right)$
(the function $\beta_{a,b}\left(\lambda\right)$ is defined in Appendix~\ref{subsec: Appendix Logarithmic Term}),
and where we also defined
\begin{equation}
E\left(\lambda\right)=\left(\lambda-1\right)^{\left(k_{-}+k_{F,R}\right)/2\pi}\left(\lambda-\nu_{0}\right)^{\Delta k/2\pi}\left(\lambda+1\right)^{1-\left(k_{F,R}+k_{+}\right)/2\pi},
\end{equation}
and
\begin{equation}
g_{j}\left(k\right)=\begin{cases}
e^{i\pi\beta_{j}} & 0\le k<k_{j},\\
e^{-i\pi\beta_{j}} & k_{j}\le k<2\pi.
\end{cases}
\end{equation}
According to the Fisher-Hartwig conjecture, the asymptotics of the
Toeplitz determinant $\tilde{D}_{L}\left(\lambda\right)$, arising
from this symbol, is given by~\citep{Jin2004,10.2307/23030524}
\begin{align}
\tilde{D}_{L}\left(\lambda\right) & \sim E\left(\lambda\right)^{L}L^{-\sum_{j=1}^{3}\beta_{j}^{2}}\nonumber \\
 & \times\prod_{1\le j<m\le3}\left|e^{ik_{j}}-e^{ik_{m}}\right|^{2\beta_{j}\beta_{m}}\prod_{j=1}^{3}G\left(1+\beta_{j}\right)G\left(1-\beta_{j}\right),
\end{align}
where $G\left(x\right)$ is the Barnes G-function~\citep{NIST:DLMF},
which obeys in particular
\begin{equation}
G\left(1+z\right)=\Gamma\left(z\right)G\left(z\right).\label{eq: Barnes function property}
\end{equation}
The required subleading term $\tilde{{\cal I}}_{{\rm const}}\left(n,\alpha\right)$
in Eq.~(\ref{eq: Entropy FH asymptotics}) is therefore given by
the integral
\begin{equation}
\tilde{{\cal I}}_{{\rm const}}\left(n,\alpha\right)=-\lim_{\varepsilon,\delta\rightarrow0^{+}}\frac{1}{2\pi i}\underset{c\left(\varepsilon,\delta\right)}{\int}\ln\left[\frac{\tilde{D}_{L}\left(\lambda\right)}{E\left(\lambda\right)^{L}L^{-\sum_{j=1}^{3}\beta_{j}^{2}}}\right]\frac{d}{d\lambda}e_{n}^{\left(\alpha\right)}\left(1+\varepsilon,\lambda\right)d\lambda.
\end{equation}

Let us now denote
\begin{equation}
\omega_{jm}=-\lim_{\varepsilon,\delta\rightarrow0^{+}}\frac{1}{2\pi i}\underset{c\left(\varepsilon,\delta\right)}{\int}\ln\left[\left|e^{ik_{j}}-e^{ik_{m}}\right|^{2\beta_{j}\beta_{m}}\right]\frac{d}{d\lambda}e_{n}^{\left(\alpha\right)}\left(1+\varepsilon,\lambda\right)d\lambda.
\end{equation}
Then, relying on the property from Eq.~$\eqref{eq: Beta on real line}$,
we obtain
\begin{align}
\omega_{12} & =-\frac{\ln\left|2\sin\left(\frac{1}{2}\Delta k\right)\right|}{2\pi^{2}}\left[\underset{-1}{\overset{\nu_{0}}{\int}}\ln\left|\frac{x-1}{x-\nu_{0}}\right|\frac{d}{dx}e_{n}^{\left(\alpha\right)}\left(1,x\right)dx+\underset{\nu_{0}}{\overset{1}{\int}}\ln\left|\frac{x-\nu_{0}}{x+1}\right|\frac{d}{dx}e_{n}^{\left(\alpha\right)}\left(1,x\right)dx\right],\nonumber \\
\omega_{13} & =\frac{\ln\left|2\sin\left(\frac{k_{-}+k_{F,R}}{2}\right)\right|}{2\pi^{2}}\left[\underset{-1}{\overset{1}{\int}}\ln\left|\frac{x-1}{x-\nu_{0}}\right|\frac{d}{dx}e_{n}^{\left(\alpha\right)}\left(1,x\right)dx+\underset{\nu_{0}}{\overset{1}{\int}}\ln\left|\frac{x-1}{x+1}\right|\frac{d}{dx}e_{n}^{\left(\alpha\right)}\left(1,x\right)dx\right],\nonumber \\
\omega_{23} & =\frac{\ln\left|2\sin\left(\frac{k_{+}+k_{F,R}}{2}\right)\right|}{2\pi^{2}}\left[\underset{-1}{\overset{\nu_{0}}{\int}}\ln\left|\frac{x-1}{x+1}\right|\frac{d}{dx}e_{n}^{\left(\alpha\right)}\left(1,x\right)dx+\underset{-1}{\overset{1}{\int}}\ln\left|\frac{x-\nu_{0}}{x+1}\right|\frac{d}{dx}e_{n}^{\left(\alpha\right)}\left(1,x\right)dx\right],
\end{align}
and thus, when we sum up the different contributions, we have
\begin{align}
\omega_{12}+\omega_{13}+\omega_{23} & =\ln\left|\frac{2\sin\left(\frac{k_{-}+k_{F,R}}{2}\right)\sin\left(\frac{1}{2}\Delta k\right)}{\sin\left(\frac{k_{+}+k_{F,R}}{2}\right)}\right|Q_{n}\left(\nu_{0},\alpha\right)\nonumber \\
 & +\ln\left|\frac{2\sin\left(\frac{k_{+}+k_{F,R}}{2}\right)\sin\left(\frac{1}{2}\Delta k\right)}{\sin\left(\frac{k_{-}+k_{F,R}}{2}\right)}\right|Q_{n}\left(-\nu_{0},-\alpha\right)\nonumber \\
 & +\ln\left|\frac{2\sin\left(k_{F,R}\right)\sin\left(k_{0}\right)}{\sin\left(\frac{1}{2}\Delta k\right)}\right|\left[\frac{1}{12}\left(\frac{1}{n}-n\right)-\frac{\alpha^{2}}{4\pi^{2}n}\right],
\end{align}
employing the notation from Eq.~(\ref{eq: Discontinuity kernel notation}).

Let us further denote
\begin{equation}
\rho_{j}=-\lim_{\varepsilon,\delta\rightarrow0^{+}}\frac{1}{2\pi i}\underset{c\left(\varepsilon,\delta\right)}{\int}\ln\left[G\left(1+\beta_{j}\right)G\left(1-\beta_{j}\right)\right]\frac{d}{d\lambda}e_{n}^{\left(\alpha\right)}\left(1+\varepsilon,\lambda\right)d\lambda,
\end{equation}
and examine, for example, $\rho_{1}$. Then, using Eq.~(\ref{eq: Beta on real line}),
we have
\begin{equation}
\rho_{1}=\frac{1}{2\pi i}\underset{\nu_{0}}{\overset{1}{\int}}\ln\left[\frac{G\left(\frac{3}{2}+\frac{1}{2\pi i}\ln\left|\frac{x-1}{x-\nu_{0}}\right|\right)G\left(\frac{1}{2}-\frac{1}{2\pi i}\ln\left|\frac{x-1}{x-\nu_{0}}\right|\right)}{G\left(\frac{1}{2}+\frac{1}{2\pi i}\ln\left|\frac{x-1}{x-\nu_{0}}\right|\right)G\left(\frac{3}{2}-\frac{1}{2\pi i}\ln\left|\frac{x-1}{x-\nu_{0}}\right|\right)}\right]\frac{d}{dx}e_{n}^{\left(\alpha\right)}\left(1,x\right)dx,
\end{equation}
and by applying Eq.~(\ref{eq: Barnes function property}) we obtain
$\rho_{1}=\Upsilon_{n}\left(\nu_{0},\alpha\right)$, per the notation
in Eq.~(\ref{eq: Upsilon def}). In a similar manner we find that
$\rho_{2}=\Upsilon_{n}\left(-\nu_{0},-\alpha\right)$ and $\rho_{3}=\Upsilon_{n}\left(-1,\alpha\right)$.
Finally, since
\begin{equation}
\tilde{{\cal I}}_{{\rm const}}\left(n,\alpha\right)=\sum_{1\le j<m\le3}\omega_{jm}+\sum_{j=1}^{3}\rho_{j},
\end{equation}
we arrive at the desired result in Eq.~(\ref{eq: FH constant term}).

\subsection{Expansion of the generating function for $n=1$ \label{subsec: Expansion of generating function}}

Here we explicitly calculate the terms of order $\alpha$ and $\alpha^{2}$
in the power series expansion of $\ln Z_{1}\left(\alpha\right)$.
We first note that
\begin{equation}
e_{1}^{\left(\alpha\right)}\left(1,x\right)=\frac{i}{2}\left(1+x\right)\alpha+\frac{x^{2}-1}{8}\alpha^{2}+{\cal O}\left(\alpha^{3}\right),\label{eq: Kernel alpha expansion}
\end{equation}
and therefore the linear term in Eq.~(\ref{eq: Conjectured entropy asymptotics})
obeys
\begin{equation}
{\cal I}_{{\rm lin}}\left(1,\alpha\right)=\frac{1}{2\pi}\left[i\alpha\left(k_{0}+k_{F,R}\right)+\underset{k_{-}}{\overset{k_{+}}{\int}}\left(\frac{i\alpha}{2}\nu\left(k\right)+\frac{\nu\left(k\right)^{2}-1}{8}\alpha^{2}\right)dk\right]+{\cal O}\left(\alpha^{3}\right).
\end{equation}

For the logarithmic and constant terms we use the fact that $Q_{1}\left(\nu,\alpha\right)$
may be calculated explicitly for any $-1\le\nu\le1$. Indeed, the
change of variables $u=\frac{x-\nu}{1-x}$ allows us to write
\begin{equation}
Q_{1}\left(\nu,\alpha\right)=-\frac{1-\nu}{2\pi^{2}}i\sin\frac{\alpha}{2}\underset{0}{\overset{\infty}{\int}}\frac{\ln u}{\left(1+u\right)\cos\frac{\alpha}{2}+i\left(u+\nu\right)\sin\frac{\alpha}{2}}\cdot\frac{du}{1+u},
\end{equation}
and then solve the integral using complex contour integration of the
function $\ln^{2}z/\left[\left(1+z\right)\cos\frac{\alpha}{2}+i\left(z+\nu\right)\sin\frac{\alpha}{2}\right]\left(1+z\right)$.
We eventually obtain 
\begin{equation}
Q_{1}\left(\nu,\alpha\right)=\frac{\ln^{2}\left[e^{-i\alpha/2}\left(\cos\frac{\alpha}{2}+i\nu\sin\frac{\alpha}{2}\right)\right]}{4\pi^{2}},
\end{equation}
\sloppy where the logarithm should be interpreted as belonging to
the principal branch, $\left|{\rm Im}\ln z\right|<\pi$. Expanding
in powers of $\alpha$, we arrive at
\begin{equation}
Q_{1}\left(\nu,\alpha\right)=-\frac{\left(1-\nu\right)^{2}}{16\pi^{2}}\alpha^{2}+{\cal O}\left(\alpha^{3}\right).
\end{equation}

Finally, we estimate the contributions of terms of the form $\Upsilon_{1}\left(\nu,\alpha\right)$
that appear in the expression for the constant term in Eq.~(\ref{eq: FH constant term}).
Up to order $\alpha^{2}$ we have, according to Eqs.~(\ref{eq: Upsilon def})
and~(\ref{eq: Kernel alpha expansion}),
\begin{equation}
\Upsilon_{1}\left(\nu,\alpha\right)=\frac{1}{2\pi i}\underset{\nu}{\overset{1}{\int}}\ln\frac{\Gamma\left(\frac{1}{2}+\frac{1}{2\pi i}\ln\left(\frac{1-x}{x-\nu}\right)\right)}{\Gamma\left(\frac{1}{2}-\frac{1}{2\pi i}\ln\left(\frac{1-x}{x-\nu}\right)\right)}\left[\frac{i}{2}\alpha+\frac{x}{4}\alpha^{2}\right]dx+{\cal O}\left(\alpha^{3}\right).
\end{equation}
Employing a change of variables $\zeta=\ln\left(\frac{x-\nu}{1-x}\right)$
and the useful formula~\citep{Jin2004}
\begin{equation}
\ln\frac{\Gamma\left(\frac{1}{2}-iw\right)}{\Gamma\left(\frac{1}{2}+iw\right)}=-i\underset{0}{\overset{\infty}{\int}}\left[2we^{-t}-\frac{\sin\left(wt\right)}{\sinh\left(t/2\right)}\right]\frac{dt}{t},\label{eq: logGamma integral}
\end{equation}
one ends up with
\begin{equation}
\Upsilon_{1}\left(\nu,\alpha\right)=\frac{1-\nu}{8\pi}\underset{-\infty}{\overset{\infty}{\int}}d\zeta\frac{e^{-\zeta}}{\left(1+e^{-\zeta}\right)^{2}}\left(2i\alpha+\frac{1+\nu e^{-\zeta}}{1+e^{-\zeta}}\alpha^{2}\right)\underset{0}{\overset{\infty}{\int}}\frac{dt}{t}\left[\frac{\zeta}{\pi}e^{-t}-\frac{\sin\left(\frac{\zeta t}{2\pi}\right)}{\sinh\left(t/2\right)}\right]+{\cal O}\left(\alpha^{3}\right).
\end{equation}
Switching the order of integration, the ${\cal O}\left(\alpha\right)$
term vanishes trivially due to the integrand being odd with respect
to $\zeta$. Carrying out the integration of the rest, we arrive at
\begin{equation}
\Upsilon_{1}\left(\nu,\alpha\right)=-\frac{\left(1-\nu\right)^{2}}{16\pi^{2}}\left(1+\gamma_{E}\right)\alpha^{2}+{\cal O}\left(\alpha^{3}\right),
\end{equation}
where $\gamma_{E}=\underset{0}{\overset{\infty}{\int}}\frac{e^{-t}+t-1}{t\left(e^{t}-1\right)}dt$
is the Euler-Mascheroni constant~\citep{whittaker_watson_1996}.

Adding up the different terms, we conclude that the expansion up to
order $\alpha^{2}$ of the nonequilibrium deviation of the generating
function (Eq.~(\ref{eq:genfun-nonequilibrium-deviation})) for $n=1$
is given by 
\begin{align}
\ln\frac{Z_{1}\left(\alpha\right)}{Z_{1}^{{\rm eq}}\left(\alpha\right)} & \approx\frac{i}{2\pi}\left[k_{F,R}-k_{0}+\frac{1}{2}\underset{k_{-}}{\overset{k_{+}}{\int}}\nu\left(k\right)dk\right]\alpha L-\left(\underset{k_{-}}{\overset{k_{+}}{\int}}\frac{1-\nu\left(k\right)^{2}}{16\pi}dk\right)\alpha^{2}L\nonumber \\
 & +\left(1-\left(\frac{1-\nu\left(k_{-}\right)}{2}\right)^{2}-\left(\frac{1+\nu\left(k_{+}\right)}{2}\right)^{2}\right)\frac{\alpha^{2}\ln L}{4\pi^{2}}\nonumber \\
 & +\frac{1-\nu_{0}^{2}}{8\pi^{2}}\left(1+\gamma_{E}+\ln\left|2\sin\frac{\Delta k}{2}\right|\right)\alpha^{2}\nonumber \\
 & -\left[\frac{1-\nu_{0}}{4\pi^{2}}\ln\left|\frac{\sin\frac{k_{-}+k_{F,R}}{2}}{\sin k_{0}}\right|+\frac{1+\nu_{0}}{4\pi^{2}}\ln\left|\frac{\sin\frac{k_{+}+k_{F,R}}{2}}{\sin k_{0}}\right|\right]\alpha^{2}+{\cal O}\left(\alpha^{3}\right).
\end{align}
Applying Eq.~(\ref{eq: definition of nu}), the expansion may be
also expressed as
\begin{align}
\ln\frac{Z_{1}\left(\alpha\right)}{Z_{1}^{{\rm eq}}\left(\alpha\right)} & \approx-\frac{i}{2\pi}\left[\underset{k_{F,R}}{\overset{k_{F,L}}{\int}}\left|r_{R}\left(k\right)\right|^{2}dk\right]\alpha L-\left(\underset{k_{-}}{\overset{k_{+}}{\int}}\frac{\left|t_{L}\left(k\right)r_{R}\left(k\right)\right|^{2}}{4\pi}dk\right)\alpha^{2}L\nonumber \\
 & +\frac{1}{4\pi^{2}}\left(1-\left|r_{R}\left(k_{F,R}\right)\right|^{4}-\left|t_{L}\left(k_{F,L}\right)\right|^{4}\right)\alpha^{2}\ln L\nonumber \\
 & +\frac{\left|t_{L}\left(k_{0}\right)r_{R}\left(k_{0}\right)\right|^{2}}{2\pi^{2}}\left(1+\gamma_{E}+\ln\left|2\sin\frac{\Delta k}{2}\right|\right)\alpha^{2}\nonumber \\
 & -\frac{\left|r_{R}\left(k_{0}\right)\right|^{2}}{2\pi^{2}}\ln\left|\frac{\sin k_{F,R}}{\sin k_{0}}\right|\alpha^{2}+{\cal O}\left(\alpha^{3}\right).
\end{align}

\subsection{The von-Neumman entanglement entropy\label{subsec: The von-Neumman-entanglement}}

We present here details of the derivation of the asymptotic form for
the unresolved vNEE in Eq.~(\ref{eq: Unresolved vNEE asymptotics}).
From Eqs.~(\ref{eq: vNEE relation to Renyi moments}) and~(\ref{eq: Generating function asymptotic expression}),
along with the fact that $Z_{1}=1$ by definition, we draw the following
relation:
\begin{equation}
{\cal C}_{{\rm lin}}L+{\cal C}_{{\rm log}}\ln L+{\cal C}_{{\rm const}}=-\lim_{n\to1}\left[\partial_{n}{\cal I}_{{\rm lin}}\left(n,0\right)L+\partial_{n}{\cal I}_{{\rm log}}\left(n,0\right)\ln L+\partial_{n}\tilde{{\cal I}}_{{\rm const}}\left(n,0\right)\right].
\end{equation}
The explicit expression for ${\cal C}_{{\rm lin}}$ in Eq.~(\ref{eq: vNEE linear coefficient})
is then straightforward to obtain. Calculating analytically the derivatives
of $\partial_{n}{\cal I}_{{\rm log}}\left(n,0\right)$ and $\partial_{n}\tilde{{\cal I}}_{{\rm const}}\left(n,0\right)$,
however, requires a more subtle analysis. The functions $Q_{n}\left(\nu,0\right)$
and $\Upsilon_{n}\left(\nu,0\right)$ that appear in those terms (defined
in Eqs.~(\ref{eq: Discontinuity kernel notation}) and~(\ref{eq: Upsilon def}))
have integral definitions with integrands that depend on $n$, and
the integrals must be rewritten before one can estimate the required
derivatives simply by taking the derivatives of the integrands.

Both $Q_{n}\left(\nu,0\right)$ and $\Upsilon_{n}\left(\nu,0\right)$
are defined as integrals over the interval $\left[\nu,1\right]$.
By splitting it into the intervals $\left[\nu,\frac{1+\nu}{2}\right]$
and $\left[\frac{1+\nu}{2},1\right]$, and by changing variables to
$u=\frac{x-\nu}{1-x}$ within the former and $u=\frac{1-x}{x-\nu}$
within the latter, we arrive at the following expressions:
\begin{equation}
Q_{n}\left(\nu,0\right)=\overset{1}{\underset{0}{\int}}\frac{dx}{2\pi^{2}x}\left\{ \ln\left[\left(1+\frac{1+\nu}{2}x\right)^{n}+\left(\frac{1-\nu}{2}x\right)^{n}\right]+\ln\left[\frac{\left(x+\frac{1+\nu}{2}\right)^{n}+\left(\frac{1-\nu}{2}\right)^{n}}{\left(\frac{1+\nu}{2}\right)^{n}+\left(\frac{1-\nu}{2}\right)^{n}}\right]\right\} -\frac{n}{12},\label{eq: Discontinuity kernel rewritten}
\end{equation}
and
\begin{align}
\Upsilon_{n}\left(\nu,0\right) & =\overset{1}{\underset{0}{\int}}\frac{dx}{2\pi^{2}x}\left\{ \ln\left[\left(1+\frac{1+\nu}{2}x\right)^{n}+\left(\frac{1-\nu}{2}x\right)^{n}\right]+\ln\left[\frac{\left(x+\frac{1+\nu}{2}\right)^{n}+\left(\frac{1-\nu}{2}\right)^{n}}{\left(\frac{1+\nu}{2}\right)^{n}+\left(\frac{1-\nu}{2}\right)^{n}}\right]\right\} \nonumber \\
 & \times\underset{0}{\overset{\infty}{\int}}\left[\frac{\cos\left(\frac{\ln x}{2\pi}z\right)}{2\sinh\left(\frac{z}{2}\right)}-\frac{e^{-z}}{z}\right]dz-n\kappa_{0},\label{eq: Upsilon rewritten}
\end{align}
where the numerical constant $\kappa_{0}$ has been defined right
after Eq.~(\ref{eq: Auxiliary functions vNEE}). In the derivation
of Eq.~(\ref{eq: Upsilon rewritten}), the formula of Eq.~(\ref{eq: logGamma integral})
was employed. The dependence on $n$ of the integrals featured in
Eqs.~(\ref{eq: Discontinuity kernel rewritten}) and~(\ref{eq: Upsilon rewritten})
is manifested in terms of the same form, the corresponding derivative
of which is given by
\begin{align}
 & \lim_{n\to1}\partial_{n}\left\{ \ln\left[\left(1+\frac{1+\nu}{2}x\right)^{n}+\left(\frac{1-\nu}{2}x\right)^{n}\right]+\ln\left[\frac{\left(x+\frac{1+\nu}{2}\right)^{n}+\left(\frac{1-\nu}{2}\right)^{n}}{\left(\frac{1+\nu}{2}\right)^{n}+\left(\frac{1-\nu}{2}\right)^{n}}\right]\right\} \nonumber \\
 & =\frac{\left(1+\frac{1+\nu}{2}x\right)\ln\left(1+\frac{1+\nu}{2}x\right)+\frac{1-\nu}{2}x\ln x+\left(x+\frac{1+\nu}{2}\right)\ln\left(x+\frac{1+\nu}{2}\right)}{1+x}-\left(\frac{1+\nu}{2}\right)\ln\left(\frac{1+\nu}{2}\right).
\end{align}
Finally, by recalling Eq.~(\ref{eq: definition of nu}), we reach
the results in Eqs.~(\ref{eq: vNEE log coefficient}) and~(\ref{eq: vNEE const coefficient}).

\subsection{Correlation matrix in a subsystem between two scatterers\label{subsec:Correlation-matrix-two-scatterers}}

We derive the approximate form appearing in Eqs.~(\ref{eq: Correlation matrix between two scatterers})
and~(\ref{eq: Toeplitz symbol between two scatterers}) for the correlation
matrix of a subsystem situated between two scattering regions --
region ${\rm I}$ on its left and region ${\rm II}$ on its right
-- assuming that the edges of the subsystem are far away from both
regions. Energy eigenstates are given by scattering states, which
for a wave incoming from the left take the form
\begin{equation}
\langle n|\psi_{k}^{\left(L\right)}\rangle=\frac{1}{\sqrt{N}}\cdot\frac{t_{L}^{\left({\rm I}\right)}\left(k\right)}{1-e^{2ik\ell}r_{R}^{\left({\rm I}\right)}\left(k\right)r_{L}^{\left({\rm II}\right)}\left(k\right)}\left[e^{ikn}+e^{2ik\ell}r_{L}^{\left({\rm II}\right)}\left(k\right)e^{-ikn}\right],\label{eq: Two scatterers left incoming wave}
\end{equation}
while for a wave incoming from the right,
\begin{equation}
\langle n|\psi_{k}^{\left(R\right)}\rangle=\frac{1}{\sqrt{N}}\cdot\frac{t_{R}^{\left({\rm II}\right)}\left(k\right)}{1-e^{2ik\ell}r_{R}^{\left({\rm I}\right)}\left(k\right)r_{L}^{\left({\rm II}\right)}\left(k\right)}\left[r_{R}^{\left({\rm I}\right)}\left(k\right)e^{ikn}+e^{-ikn}\right],\label{eq: Two scatterers right incoming wave}
\end{equation}
where $\ell$ is the number of sites between the scattering regions,
and the convention $k>0$ is used. By neglecting the effect of localized
bound states as in Sec.~\ref{sec: The Physical Model}, we may use
Eqs.~(\ref{eq: Two scatterers left incoming wave}) and~(\ref{eq: Two scatterers right incoming wave})
in order to write the creation operator associated with site $n$
as
\begin{align}
a_{n}^{\dagger} & =\sum_{k>0}\frac{1}{\sqrt{N}}\left(\frac{t_{L}^{\left({\rm I}\right)}\left(k\right)}{1-e^{2ik\ell}r_{R}^{\left({\rm I}\right)}\left(k\right)r_{L}^{\left({\rm II}\right)}\left(k\right)}\right)^{*}\left[e^{-ikn}+e^{-2ik\ell}r_{L}^{\left({\rm II}\right)}\left(k\right)^{*}e^{ikn}\right]a_{k,L}^{\dagger}\nonumber \\
 & +\sum_{k>0}\frac{1}{\sqrt{N}}\left(\frac{t_{R}^{\left({\rm II}\right)}\left(k\right)}{1-e^{2ik\ell}r_{R}^{\left({\rm I}\right)}\left(k\right)r_{L}^{\left({\rm II}\right)}\left(k\right)}\right)^{*}\left[r_{R}^{\left({\rm I}\right)}\left(k\right)^{*}e^{-ikn}+e^{ikn}\right]a_{k,R}^{\dagger}.
\end{align}
By exchanging summation with integration and neglecting all Hankel
terms, the elements of the correlation matrix can now be written as
\begin{align}
C_{mn} & \approx\underset{0}{\overset{k_{F,L}}{\int}}\frac{dk}{2\pi}\left|\frac{t_{L}^{\left({\rm I}\right)}\left(k\right)}{1-e^{2ik\ell}r_{R}^{\left({\rm I}\right)}\left(k\right)r_{L}^{\left({\rm II}\right)}\left(k\right)}\right|^{2}\left[e^{-i\left(m-n\right)k}+\left|r_{L}^{\left({\rm II}\right)}\left(k\right)\right|^{2}e^{i\left(m-n\right)k}\right]\nonumber \\
 & +\underset{0}{\overset{k_{F,R}}{\int}}\frac{dk}{2\pi}\left|\frac{t_{R}^{\left({\rm II}\right)}\left(k\right)}{1-e^{2ik\ell}r_{R}^{\left({\rm I}\right)}\left(k\right)r_{L}^{\left({\rm II}\right)}\left(k\right)}\right|^{2}\left[\left|r_{R}^{\left({\rm I}\right)}\left(k\right)\right|^{2}e^{-i\left(m-n\right)k}+e^{i\left(m-n\right)k}\right].
\end{align}

Next, we employ the approximation
\begin{equation}
\frac{1}{\left|1-e^{2ik\ell}r_{R}^{\left({\rm I}\right)}\left(k\right)r_{L}^{\left({\rm II}\right)}\left(k\right)\right|^{2}}\approx\frac{1}{1-\left|r_{R}^{\left({\rm I}\right)}\left(k\right)r_{L}^{\left({\rm II}\right)}\left(k\right)\right|^{2}},
\end{equation}
which is justified given that the difference between the two expressions
oscillates as a function of $k$ with a frequency of $2\ell$, and
thus after integration its contribution decays for $\ell$ that is
large with respect to the Fermi wavelengths, $\left(k_{F,L}\right)^{-1}$
and $\left(k_{F,R}\right)^{-1}$. We therefore obtain
\begin{align}
C_{mn} & \approx\underset{-k_{F,R}}{\overset{-k_{F,L}}{\int}}\frac{dk}{2\pi}\cdot\frac{\left|t_{R}^{\left({\rm II}\right)}\left(-k\right)\right|^{2}}{1-\left|r_{R}^{\left({\rm I}\right)}\left(-k\right)r_{L}^{\left({\rm II}\right)}\left(-k\right)\right|^{2}}e^{-i\left(m-n\right)k}\nonumber \\
 & +\underset{-k_{F,L}}{\overset{k_{F,R}}{\int}}\frac{dk}{2\pi}e^{-i\left(m-n\right)k}+\underset{k_{F,R}}{\overset{k_{F,L}}{\int}}\frac{dk}{2\pi}\cdot\frac{\left|t_{L}^{\left({\rm I}\right)}\left(k\right)\right|^{2}}{1-\left|r_{R}^{\left({\rm I}\right)}\left(k\right)r_{L}^{\left({\rm II}\right)}\left(k\right)\right|^{2}}e^{-i\left(m-n\right)k}.
\end{align}
This result holds true regardless of the direction of the bias voltage.
Once we explicitly distinguish between the two possible directions,
we finally obtain the expression in Eqs.~(\ref{eq: Correlation matrix between two scatterers})
and~(\ref{eq: Toeplitz symbol between two scatterers}).

\section{Additional plots for the single impurity model\label{sec:Appendix:-Additional-plots}}

\begin{figure}
\begin{centering}
\includegraphics[viewport=90bp 15bp 1250bp 565bp,clip,scale=0.35]{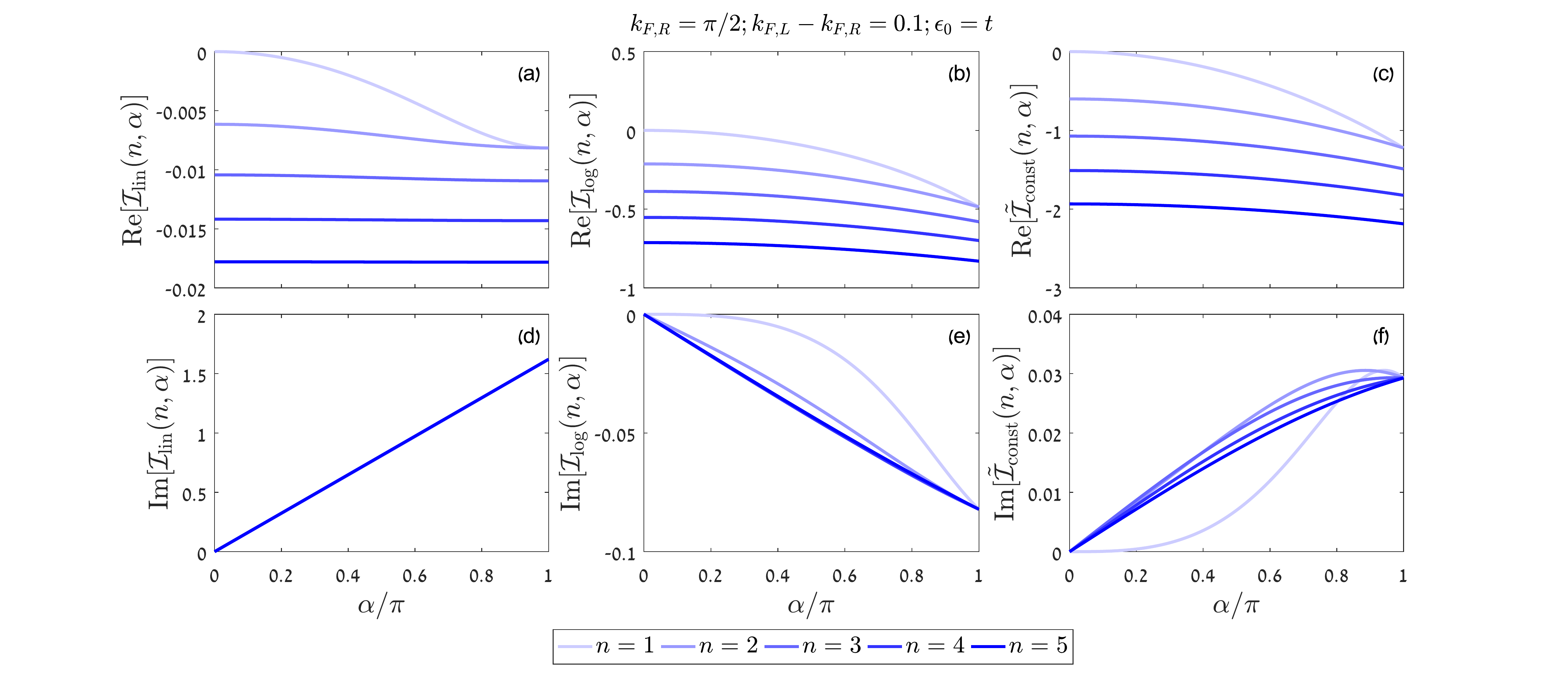}
\par\end{centering}
\caption{\label{fig: Coefficients alpha dependence}The single impurity model:
Coefficients in the analytical asymptotic expression for $\ln Z_{n}\left(\alpha\right)$
in Eq.~(\ref{eq: Generating function asymptotic expression}), as
a function of $\alpha$ for different fixed values of $n$. The remaining
parameters are set to $\epsilon_{0}=t$, $k_{F,R}=\pi/2$ and $k_{F,L}-k_{F,R}=0.1$.
The top panels (a)--(c) show the real parts of the coefficients,
and the bottom panels (d)--(f) show their imaginary parts. Panels
(a) and (d) are for the coefficient ${\cal I}_{{\rm lin}}$, panels
(b) and (e) are for ${\cal I}_{{\rm log}}$, and panels (c) and (f)
are for $\tilde{{\cal I}}_{{\rm const}}$. Note that in panel (d)
the different curves overlap because ${\rm Im}\left[{\cal I}_{{\rm lin}}\right]$
is dominated by the first term in Eq.~(\ref{eq: Conjectured entropy asymptotics}),
which is independent of $n$.}
\end{figure}

\begin{figure}[H]
\begin{centering}
\includegraphics[viewport=90bp 15bp 1250bp 565bp,clip,scale=0.35]{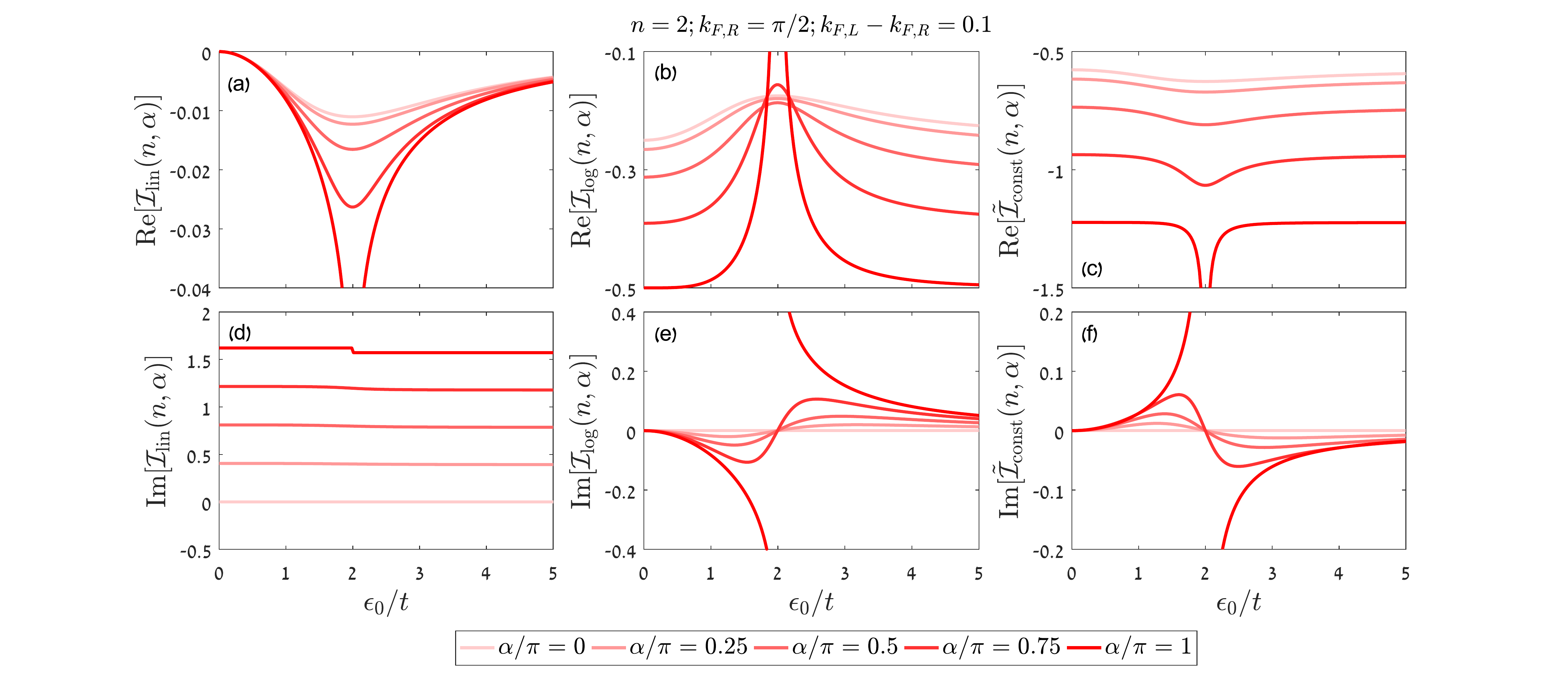}
\par\end{centering}
\caption{\label{fig: Coefficients e_0 dependence}The single impurity model:
Coefficients in the analytical asymptotic expression for $\ln Z_{n}\left(\alpha\right)$
in Eq.~(\ref{eq: Generating function asymptotic expression}), as
a function of $\epsilon_{0}/t$ for different fixed values of $\alpha$.
The remaining parameters are set to $n=2$, $k_{F,R}=\pi/2$ and $k_{F,L}-k_{F,R}=0.1$.
The top panels (a)--(c) show the real parts of the coefficients,
and the bottom panels (d)--(f) show their imaginary parts. Panels
(a) and (d) are for the coefficient ${\cal I}_{{\rm lin}}$, panels
(b) and (e) are for ${\cal I}_{{\rm log}}$, and panels (c) and (f)
are for $\tilde{{\cal I}}_{{\rm const}}$.}
\end{figure}

\begin{figure}[H]
\begin{centering}
\includegraphics[viewport=90bp 15bp 1250bp 565bp,clip,scale=0.35]{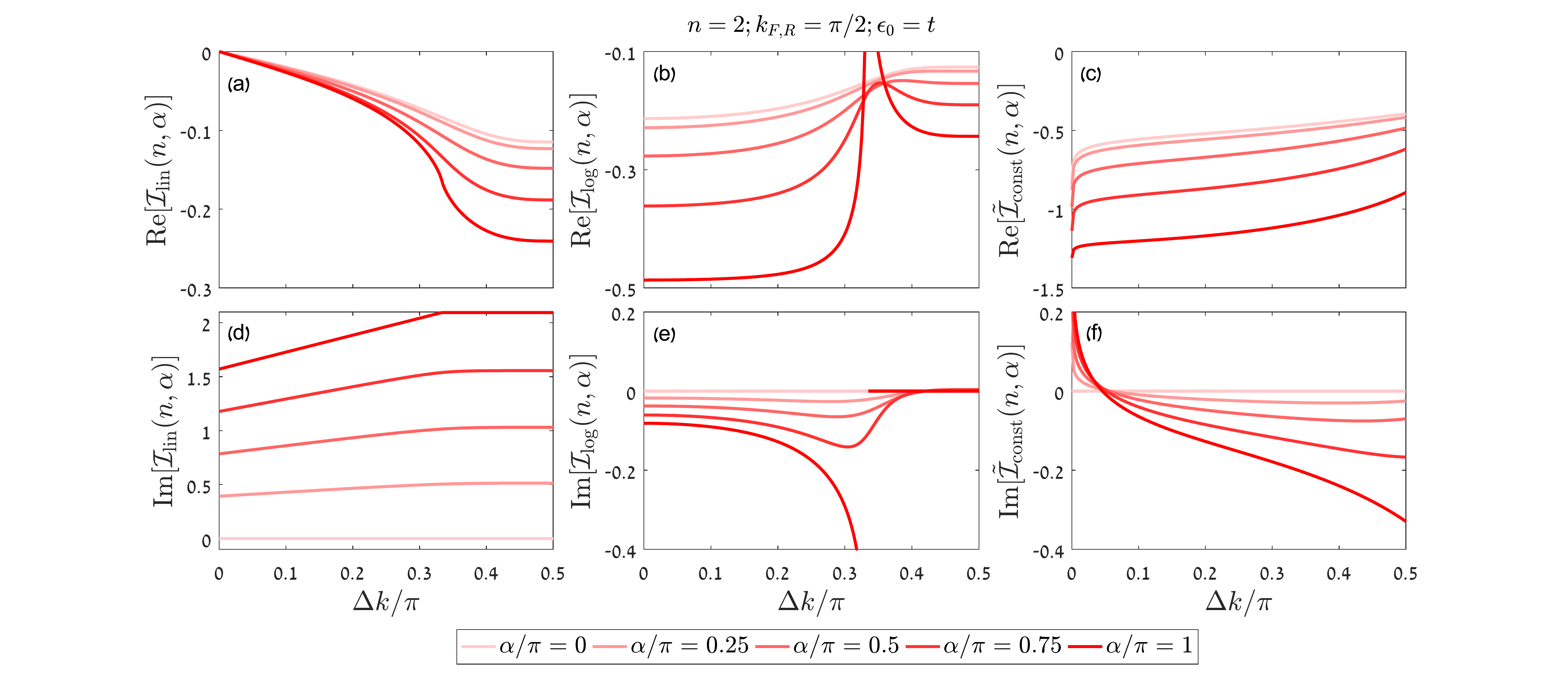}
\par\end{centering}
\caption{\label{fig: Coefficients dk dependence}The single impurity model:
Coefficients in the analytical asymptotic expression for $\ln Z_{n}\left(\alpha\right)$
in Eq.~(\ref{eq: Generating function asymptotic expression}), as
a function of $\Delta k=k_{F,L}-k_{F,R}$ for different fixed values
of $\alpha$. The remaining parameters are set to $n=2$, $\epsilon_{0}=t$
and $k_{F,R}=\pi/2$. The top panels (a)--(c) show the real parts
of the coefficients, and the bottom panels (d)--(f) show their imaginary
parts. Panels (a) and (d) are for the coefficient ${\cal I}_{{\rm lin}}$,
panels (b) and (e) are for ${\cal I}_{{\rm log}}$, and panels (c)
and (f) are for $\tilde{{\cal I}}_{{\rm const}}$.}
\end{figure}

Focusing on the single impurity model defined in Sec.~\ref{sec: The-Single-Impurity},
the plots presented here illustrate the parameter dependence of the
coefficients in the analytical asymptotics of the generating function
(Eq.~(\ref{eq: Generating function asymptotic expression})). Fig.~\ref{fig: Coefficients alpha dependence}
demonstrates how, as $n$ increases, the dependence of ${\rm Re}\ln Z_{n}\left(\alpha\right)$
on $\alpha$ becomes flatter, which amounts to a narrower distribution
of the corresponding charge-resolved R\'enyi moment about its peak.
Figs.~\ref{fig: Coefficients e_0 dependence}--\ref{fig: Coefficients dk dependence}
depict the dependence of the coefficients on $\epsilon_{0}/t$ and
$k_{\pm}$ (respectively), where the most conspicuous features are
related to the divergences of the coefficients ${\cal I}_{{\rm log}}$
and $\tilde{{\cal I}}_{{\rm const}}$ that are discussed at length
in Subsec.~\ref{subsec:Subleading-asymptotics-of}. In Figs.~\ref{fig: Coefficients e_0 dependence}(b),(e)
and~\ref{fig: Coefficients dk dependence}(b),(e) the divergences
of ${\cal I}_{{\rm log}}$ for $\alpha=\pi$ are related to points
where either $\nu\left(k_{-}\right)=0$ or $\nu\left(k_{+}\right)=0$.
In Figs.~\ref{fig: Coefficients e_0 dependence}(c),(f) the coefficient
$\tilde{{\cal I}}_{{\rm const}}$ is seen to diverge for $\alpha=\pi$
and $\nu_{0}=0$. Note that in Fig.~\ref{fig: Coefficients dk dependence},
the values of ${\cal I}_{{\rm log}}$ and $\tilde{{\cal I}}_{{\rm const}}$
at the limit $\Delta k\to0$ are fictitious since, as explained in
Subsec.~\ref{subsec:Subleading-asymptotics-of}, throughout our calculations
we implicitly assume $\Delta k\gg1/L$.

\bibliographystyle{SciPost_bibstyle}
\bibliography{EE_NESS}

\end{document}